\documentclass[aps,prb,twocolumn,showpacs,superscriptaddress,10pt]{revtex4-1}
\usepackage{graphicx}
\usepackage{bm}% bold math
\usepackage{amssymb}
\usepackage{amsmath}
\usepackage{amsthm}
\usepackage[utf8]{inputenc}
\usepackage{array}

\newcommand{\rmd}{\mathrm d}

\newcolumntype{L}[1]{>{\raggedright\let\newline\\\arraybackslash\hspace{0pt}}m{#1}}
\newcolumntype{C}[1]{>{\centering\let\newline\\\arraybackslash\hspace{0pt}}m{#1}}
\newcolumntype{R}[1]{>{\raggedleft\let\newline\\\arraybackslash\hspace{0pt}}m{#1}}

\begin{document}
\title{The role of weakest links and system size scaling in multiscale modeling of stochastic plasticity}

\author{P\'eter Dus\'an Isp\'anovity}
\email{ispanovity@metal.elte.hu}
\affiliation{Department of Materials Physics, E\"otv\"os University, P\'azm\'any P\'eter s\'et\'any 1/a, H-1117 Budapest, Hungary}

\author{D\'aniel T\"uzes}
\affiliation{Department of Materials Physics, E\"otv\"os University, P\'azm\'any P\'eter s\'et\'any 1/a, H-1117 Budapest, Hungary}
\affiliation{Institute for Materials Simulation (WW8), Friedrich-Alexander-University, Erlangen-N\"urnberg, Dr.-Mack-Str.~77, 
D-90762 F\"urth, Germany}

\author{P\'eter Szab\'o}
\affiliation{Department of Materials Physics, E\"otv\"os University, P\'azm\'any P\'eter s\'et\'any 1/a, H-1117 Budapest, Hungary}

\author{Michael Zaiser}
\affiliation{Institute for Materials Simulation (WW8), Friedrich-Alexander-University, Erlangen-N\"urnberg, Dr.-Mack-Str.~77, 
D-90762 F\"urth, Germany}

\author{Istv\'an Groma}
\affiliation{Department of Materials Physics, E\"otv\"os University, P\'azm\'any P\'eter s\'et\'any 1/a, H-1117 Budapest, Hungary}

\begin{abstract}
Plastic deformation of crystalline and amorphous matter often involves intermittent local strain burst events. To understand the physical background of the phenomenon a minimal stochastic mesoscopic model was introduced, where microstructural details are represented by a fluctuating local yielding threshold. In the present paper, we propose a method for determining this yield stress distribution by lower scale discrete dislocation dynamics simulations and using a weakest link argument. The success of scale-linking is demonstrated on the stress-strain curves obtained by the resulting mesoscopic and the discrete dislocation models. As shown by various scaling relations they are statistically equivalent and behave identically in the thermodynamic limit. The proposed technique is expected to be applicable for different microstructures and amorphous materials, too.
\end{abstract}
\maketitle

\section{Introduction}

Crystal plasticity involves important features on multiple spatial and temporal scales ranging from atomistic processes to, e.g.,\ the grain structure of a specimen. Understanding and modeling such a rich phenomenon requires a multiscale approach, where different level models rely on inputs from lower scales. One of such steps that has been intensively studied is to bridge discrete dislocation plasticity with a higher scale continuum description \cite{groma2003spatial, groma2015scale, hochrainer2014continuum, poh2013homogenization, levkovitch2006large, sedlavcek2007continuum}. The main motivation behind these activities is that dislocations interact with long-range stress fields, so in a discrete model all mutual pair interactions between dislocations have to be taken into account leading to a time complexity that makes calculations unsolvable already for small samples. This restriction could be lifted by an appropriate continuum model, in which dislocations are smeared out in terms of continuous density fields, and one considers the dynamics of these fields in the form of continuity equations.

These descriptions filter out spatial and temporal fluctuations appearing in the form of intermittent strain bursts caused by dislocation avalanches \cite{dimiduk2006scale, zaiser2008strain}. Such fluctuations, however, often represent important physics that one may intend to take into account. For instance, in the case of micron-scale specimens the stochastic strain bursts prohibit the predicted formability, and thus represent a major challenge for material design \cite{csikor2007dislocation}. They also play an important role in size effects, that is, the increase of material strength if specimen dimensions are reduced down to the micron regime or below \cite{Uchic_2004, dimiduk2005size, uchic2009plasticity}. So, from a technical point of view it seems desirable to extend the continuum dislocation models by a stochastic component to account for avalanche dynamics.

Such a stochastic crystal plasticity model (SCPM) was proposed by Zaiser \emph{et al.}\ in two dimensions (2D), which extended the continuum models with a random term in the local yield stress of the material \cite{Zaiser2005}. This feature is meant to account for the inherent inhomogeneity of the dislocation microstructure that often appears in the form of distinctive patterns. The model is in fact a cellular automaton (CA) representation of the plastic strain field evolution. The elementary event is the local slip of a cell (achieved in practice by the motion of nearby dislocations), that induces a long-range internal stress redistribution, that may trigger further events. The local yield threshold is updated after each event to account for microstructural rearrangements that take place during plastic slip. The resulting model recovers the stochastic nature of plasticity and yields a power-law distribution for the random steps appearing on the stress-strain curves \cite{zaiser2007slip}. It also proved successful to model the quasi-periodic oscillatory behavior observed at slow deformation of micron-scale single crystalline pillars \cite{papanikolaou2012quasi}.

Interestingly, very similar mesoscopic stochastic models were already introduced earlier with a different aim, namely, to study plasticity in amorphous materials, where the dislocation-mediated deformation mechanism is absent \cite{baret2002extremal, talamali2011avalanches, budrikis2013avalanche, sandfeld2015avalanches, PhysRevLett.116.065501}. Among these models the basic assumptions are identical to those of SCPM: (i) that plastic strain accumulates in local shear transformations that generate long-range internal stress redistribution and (ii) that the material exhibits internal disorder represented by a fluctuating local yield stress. In fact, the 2D model of Roux \emph{et al.}\cite{baret2002extremal},\ apart from a few minor technical differences, is identical to the SCPM of Zaiser \emph{et al.}\cite{Zaiser2005} The reason for this is, on the one hand, that although dislocation slip is characterized by the quantum of the Burgers vector, a local strain increment is (roughly speaking) the product of the Burgers vector and the total distance traveled by the dislocations which is, therefore, not restricted to discrete values. On the other hand, the elastic stress induced by a local plastic slip event is independent of the underlying deformation mechanism, and is described by the Eshelby solution of the corresponding eigenstrain problem \cite{Eshelby1957}. In the amorphous model of Roux \emph{et al.}\cite{baret2002extremal}\ and the SCPM of Zaiser \emph{et al.}\cite{Zaiser2005}\ the free parameters of the model are the distribution of the local yield stress and the magnitude (or distribution) of a local slip event. These parameters are representing the microstructural features of the actual material, and are, of course, expected to differ for amorphous and crystalline materials.

In the present paper, we demonstrate how these parameters can be calibrated in the case of crystalline plasticity. At the lower scale we use conventional 2D discrete dislocation dynamics (DDD) models, that have been studied extensively in the literature \cite{miguel2001intermittent, ispanovity2010submicron, tsekenis2011dislocations, ovaska2015quenched, kapetanou2015stress}. Load-controlled quasi-static plastic deformation is simulated, where individual avalanches can be identified \cite{ispanovity2014avalanches}. We find that the stress value corresponding to the first avalanche follows a Weibull distribution, and the mean stress at the $i$th avalanche represent a weakest link sequence from the same distribution. This implies that plastic events are local and the subsequent avalanches are weakly correlated confirming the main assumption of the SCPM. We also provide an in-depth statistical analysis of the stress-strain curves and show by scaling relations that at not too large strains both DDD and SCPM exhibit a smooth plastic response if the system size tends to infinity. This tendency is characterized by a size exponent that, with an appropriate choice for the local strain increment in the SCPM, equals for the different models. The so configured SCPM, thus, provides stress-strain curves statistically equivalent to those obtained by DDD.

The paper is organized as follows. Dimensionless units used in the paper are introduced in Sec.~\ref{sec:units}, followed by the description of the plasticity models in Sec.~\ref{sec:model}, and the summary of the numerical results in Sec.~\ref{sec:results}. Section \ref{sec:model} presents a plasticity theory that correctly describe the numerical findings in the microplastic regime. The paper concludes with a Discussion and a Summary section.

\section{Dimensionless units}
\label{sec:units}

Infinite dislocation systems, apart from the core region not considered here, are invariant to the following re-scaling (see, e.g.,\ Ref.~\onlinecite{zaiser2014scaling}):
\begin{equation}
\bm r \to \bm r/c\text{, } \gamma \to c\gamma \text{, and } \tau \to c\tau,
\end{equation}
where $\bm r$, $\gamma$, and $\tau$ denote the spatial coordinate, the plastic shear strain, and the shear stress, respectively, and $c>0$ is an arbitrary constant. This universal feature is a simple result of the $1/r$ type (scale-free) decay of the dislocation stress fields. This property also means that in an infinite dislocation system the only length scale is the average dislocation spacing $\rho^{-1/2}$, where $\rho$ is the total dislocation density, so, naturally, this value is chosen as $c$. In the case of strain $\gamma$ and stress $\tau$ the Burgers vector $b$, shear modulus $\mu$, and the Poisson ratio $\nu$ are also required to arrive at dimensionless units denoted by $(\cdot)'$:
\begin{equation}
\bm r' = \rho^{1/2} \bm r\text{, } \gamma' = \gamma/(b\rho^{1/2}) \text{, } \tau' = \tau/ \left( \frac{\mu b}{2\pi(1-\nu)}\rho^{1/2} \right).
\end{equation}
Throughout the paper these dimensionless units will be used, and the distinguishing $(\cdot)'$ symbol will be omitted.

\section{Simulation methods}
\label{sec:sim_methods}

In this paper, for simplicity, two-dimensional (2D) models are applied. All three models introduced below have been used extensively in the literature, therefore, only their main features are summarized.

\subsection{Stochastic continuum plasticity model (SCPM)}
\label{sec:scpm}

\begin{figure}[!htbp]
\begin{center}

\includegraphics[scale=1, angle=0]{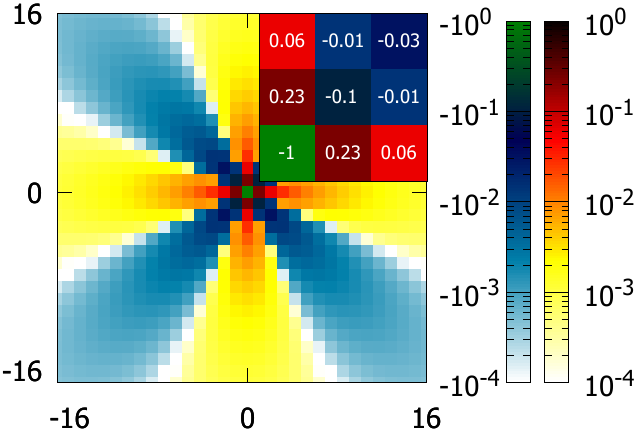}
\caption{\label{fig:stress_kernel} The center part of the stress field of an elementary slip event $\Delta {\gamma ^{{\text{pl}}}}$ for the case $L=128$ in the units of $\left| {G_{0,0}^E} \right|\Delta {\gamma ^{{\text{pl}}}}$. In the upper right corner a magnification of the cells $\left[ {0,2} \right] \times \left[ {0,2} \right]$ is shown. Note the 4-fold symmetry.
}
\end{center}
\end{figure}

The model is based on a crystal plasticity model introduced by Zaiser and Moretti \cite{Zaiser2005} and considers a plane strain problem with a local plastic shear strain field ${\gamma ^{{\text{pl}}}}\left( {\bm r} \right)$ and a local shear stress $\tau^\text{loc} \left( {\bm r} \right)$.
%If the stress acting on a local volume element exceeds a local threshold value ${\tau ^{\text{th}}}\left( \bm{r} \right)$ the plastic strain field changes at the same place. The calculation of the stress field created by such an elementary slip event is based on the point-like Eshelby inclusion problem \cite{Eshelby1957}, and may trigger further slip events leading to an avalanche. Such an event will last until the local stress gets smaller than the local threshold value everywhere in the sample.
In an infinite system one can write the stress at an arbitrary position $\bm r$ as
\begin{equation}
{\tau ^{{\text{loc}}}}\left( \bm{r} \right) = {\tau ^{{\text{ext}}}} + \left( {{G^E} * {\gamma ^{{\text{pl}}}}} \right)\left( \bm{r} \right),
\label{eq:tau_loc}
\end{equation} 
i.e.,\ it consists of two parts: an external load and internal part generated by the inhomogeneous ${\gamma ^{{\text{pl}}}}\left( {\bm r} \right)$ 
field via ${G^E}\left( {\mathbf{r}} \right)$, the elastic Green's function specified by the corresponding Eshelby inclusion problem.\cite{Eshelby1957} 
%Plastic strain can occur once the local stress exceeds the local threshold value: 
%\begin{equation}
%\label{eq:flow_rule}
%{\tau^r}\left( {{\mathbf{r}},t} \right): = {\tau ^{{\text{th}}}}\left( {{\mathbf{r}},t} \right) - \left| {{\tau ^{{\text{loc}}}}\left( {{\mathbf{r}},t} \right)} \right| \leqslant 0,
%\end{equation}
% leading to an elementary slip event. 
The stress and strain fields are discretized on a square lattice with cell size $d$ (measured in dimensionless units introduced in Sec.~\ref{sec:units}) of global size $L\cdot d \times L\cdot d$ with the edges parallel to the $x$ and $y$ direction and $L = 8, 16, ..., 8192$. The discretized Green's function $G^E_{ij}$ is proportional to the stress field of a local slip at the origin ($\gamma^\text{pl}_{ij} = \delta_{ij} \Delta \gamma^\text{pl}$) and is, therefore, calculated as %follows. The cell is cut in half along the $x$ and $y$ directions, the upper part moved along the $x$ direction and the the right part moved along the $y$ direction according to the local stress, then the formation is glued back together and elastically fitted back to its original position generating the stress field of the inclusion. This method is equivalent to
the stress field of four edge dislocations with Burgers vectors $b{{\bm{e}}_x}$, $b{{\bm{e}}_y}$, $-b{{\bm{e}}_x}$ and $-b{{\bm{e}}_y}$ at the right, top, left and bottom sides of the cell, respectively. This corresponds to a local plastic shear of $\Delta {\gamma ^{{\text{pl}}}} = 2/d$. The stress values are evaluated at the center-points of the cells and, for example, at the origin it gives $G_{0,0}^E\Delta {\gamma ^{{\text{pl}}}} =  - 4 \Delta \gamma^\text{pl} = - 8/d$. For the rest of the cells the numerical values of $G_{ij}^E$ can be seen in the units of $\left| {G_{0,0}^E} \right|$ in Fig.~\ref{fig:stress_kernel}.

The internal structural disorder is taken into account via the fluctuating local threshold value $\tau^\text{th}$. This means that if for a given cell
\begin{equation}
\label{eq:flow_rule}
{\tau^r}\left( \bm{r},t \right): = {\tau ^{{\text{th}}}}\left( \bm{r},t \right) - \left| {\tau ^\text{loc}\left( {\bm{r},t} \right)} \right| \leqslant 0
\end{equation}
holds, then it is in equilibrium, otherwise it is active, that is, it yields. We assume that cells are large enough to neglect the correlation between the threshold values of the neighbor cells making them independent random variables. The values are chosen from a Weibull distribution with shape parameter $\nu  = 1,1.4{\text{ and }}2$ and scale parameter $\tau_w$. 

At the beginning of a simulation initial values of the local threshold values are distributed and the strain (and, thus, the local stress, too) is set to zero everywhere. Then a stress-controlled loading procedure is implemented as follows. The external stress is increased until Eq.~(\ref{eq:flow_rule}) is violated in a single cell, which then becomes active. At this cell the local strain is increased by $\Delta {\gamma ^{{\text{pl}}}}$ and a new local threshold value is assigned from the threshold distribution. The internal stress is recalculated according to Eq.~(\ref{eq:tau_loc}) and the newly activated cells are determined using Eq.~(\ref{eq:flow_rule}). As long as ${\tau^r} \leqslant 0$ holds for at least one cell the system is in an avalanche, and the local strain is increased by $\Delta \gamma^\text{pl}$ in the cell where $\tau^r$ is the smallest (extremal dynamics). When for all the cells ${\tau^r} > 0$ holds the avalanche ceases and the external stress is further increased by the smallest ${\tau^r}$ to trigger the next avalanche. At every state of the system the total strain is the spatial average of the local strain: $\gamma := \langle \gamma^\text{pl} \rangle$.

\subsection{Discrete dislocation dynamics}

\subsubsection{Continuous representation (TCDDD)}
\label{sec:taddd}

The model called time-continuous discrete dislocation dynamics (TCDDD) considered here consists of $N$ straight parallel edge dislocations, all of which lay in the same slip system. The slip direction was chosen to be parallel to the $x$ side of the square-shaped simulation area, so the Burgers vectors of the dislocations, assuming that their magnitude is $b$, may point in two directions described by their \emph{sign} $s$: $\bm b_i = s_i(b, 0)$, where $i=1,\dots,N$. Since there is only one slip direction present, the interaction of the dislocations that influences glide can be described in terms of the shear stress field $\tau_{\mathrm{ind}}$ induced by each dislocation. Its form in the dimensionless units introduced above is
\begin{equation}
 \label{eq:tau-ind}
 \tau_{\mathrm{ind}}(\bm r)= x(x^2-y^2) / (x^2+y^2)^{2} = r^{-1}\cos\varphi\cos 2 \varphi, 
\end{equation}
where $\bm r=(x,y)$ is the relative displacement from the dislocation and $(r, \varphi)$ are the corresponding polar coordinates. To model an infinite crystal periodic boundary conditions were applied and the periodic form of Eq.~(\ref{eq:tau-ind}) was used (for details see, e.g.~Ref.~\onlinecite{Bako2006}).

This model aims to describe the easy slip regime, where dislocation glide is dominant, therefore, climb and cross-slip are neglected. The system is driven by a homogeneous external shear stress field $\tau_{\mathrm{ext}}$, so the equation of motion of the $i$th dislocation is:
\begin{equation}
 \dot{x}_i =s_i \Bigg[\sum_{j=1; j\neq i}^N s_j\tau_{\mathrm{ind}}(\bm r_i - \bm r_j)
+\tau_{\mathrm{ext}}\Bigg], \quad \dot{y}_i =0,
\label{eq:eqn_of_motion}
\end{equation}
where $\bm r_i = (x_i, y_i)$ denotes its position, and the dislocation mobility was absorbed into the time scale. Here it is assumed that due to the strong phonon drag the motion is overdamped and, thus, inertial terms can be neglected.

The simulations were started from a random arrangement of an equal number of positive and negative sign dislocations. First, Eq.~(\ref{eq:eqn_of_motion}) was solved at $\tau_\text{ext}=0$ until the system reached equilibrium. Then a quasistatic load-controlled procedure was applied, i.e., stress was increased with a fixed rate between avalanches, and was kept constant during the active periods (for details see Ref.~\onlinecite{Bako2006}). The plastic strain at time $t$ is obtained using $\gamma = \sum_{i=1}^N s_i (x_i(t) - x_i(0))$.

In the dimensionless units introduced above the linear system size is related only to the number of dislocations as $L=N^{0.5}$. The simulations were repeated for different system sizes $L=8, 11.31, 16, 22.63, 32$ on a large ensemble of statistically equivalent realizations in each case (consisting of $3000, 2000, 800, 300$, and $180$ individual runs, respectively). Very narrow dislocation dipoles were annihilated since they practically do not affect the dynamics but due to numerical reasons they slow down the simulations considerably.

\subsubsection{Cellular automaton representation (CADDD)}
\label{sec:caddd}

The cellular automaton discrete dislocation dynamics (CADDD) is very similar to the continuous method introduced above except for two important differences:
\begin{enumerate}
\item The space is discretized, meaning that dislocations move on a regular equidistant grid, and only one dislocation may be present in a cell at the same time. In the simulations performed the cell size $\delta$ was 128 times smaller than the average dislocation spacing, meaning that only every $128\times128$th cell was populated.
\item The time is also discretized, i.e.,\ the dynamics is defined by a rule that controls how to move dislocations from one cell to a neighbor cell. Here we use extremal dynamics (ED), meaning that the stress induced by the other dislocations $\tau$ [i.e., the RHS of Eq.~(\ref{eq:eqn_of_motion})] is evaluated at the left (right) border of the cell containing the dislocation. If the force $s_i \tau >0(<0)$ then a step in the right (left) direction is energetically favorable and the decrease in the stored elastic energy $\Delta E$ is proportional with $-|\tau| \delta$. In every timestep the single dislocation with the highest energy drop is moved, then the interaction stresses are recomputed. If there is no dislocation eligible to move (that is, $\Delta E > 0$ for each) then the external stress is increased until a dislocation starts to move. If two dislocations of opposite sign occupy the same cell, they are annihilated.
\end{enumerate}

As seen, the driving is similar to the quasistatic load-control of the TCDDD. The simulations are also started from a random dislocation configuration, and different system sizes are considered. It is noted, that due to the absence of a real time scale in this model there is no straightforward way to define a strain burst. The advantage of this model lies in its faster computational speed compared to TCDDD that allows much larger systems to be studied. In addition, it allows to test the role of the chosen dynamics (overdamped or extreme) in the results obtained.

\section{Numerical Results}
\label{sec:results}

\begin{figure}[!htbp]
\begin{center}
\includegraphics[scale=0.5, angle=0]{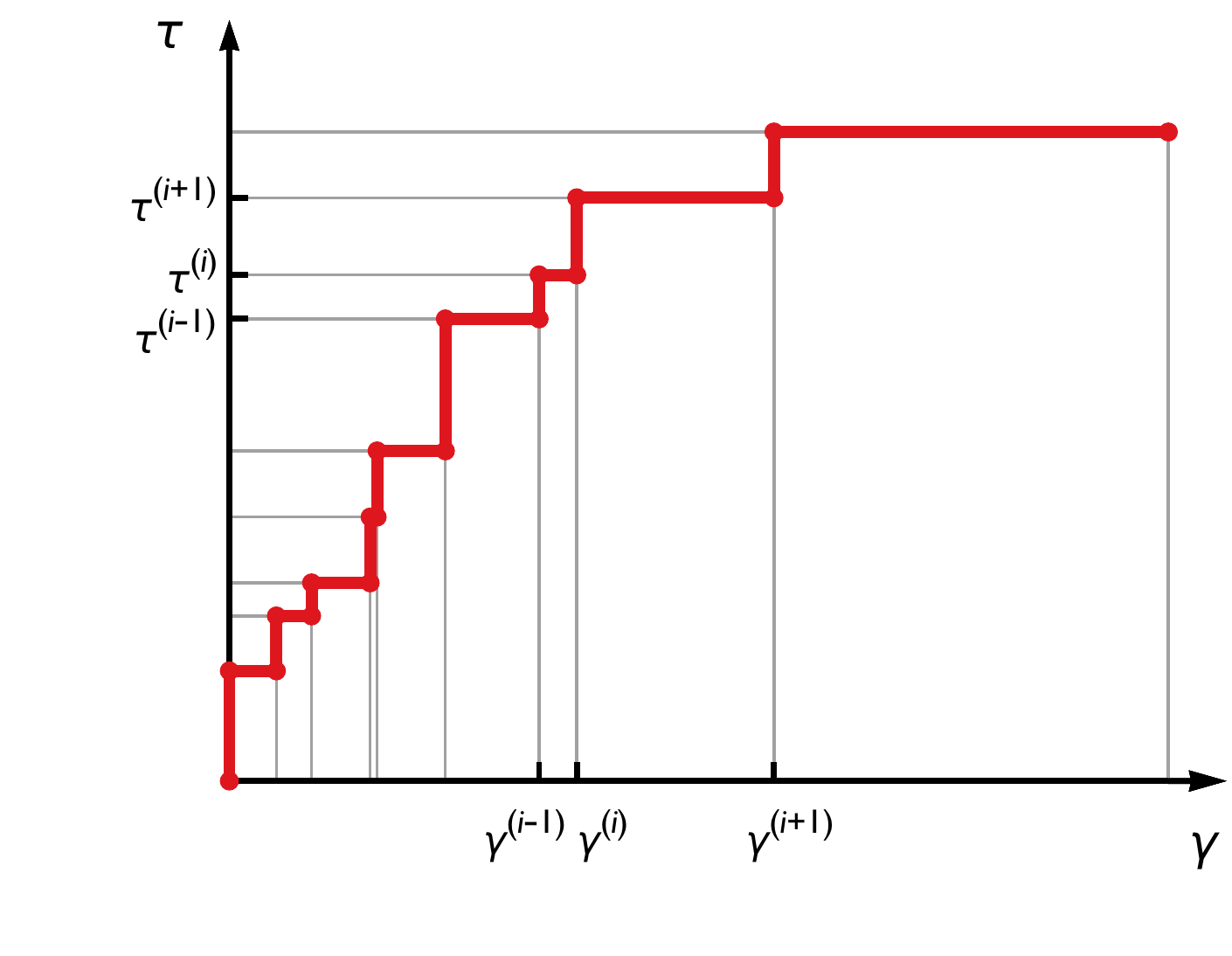}
\caption{\label{fig:stress_strain_sketch} Sketch of a stress-strain curve obtained by the models of Sec.~\ref{sec:sim_methods}. The curve can be fully characterized by the stress and strain sequence $\tau^{(i)}$ and $\gamma^{(i)}$, respectively.
}
\end{center}
\end{figure}

In this section simulation results are provided using the plasticity models introduced in Sec.~\ref{sec:sim_methods}. In every model, the obtained stress-strain curves are step-like and different for all realizations. In the following, the statistical properties of the stress-strain curves will be examined followed by the analysis of the stress and strain sequences corresponding to individual strain bursts. The latter ones (denoted by $\tau^{(i)}$ and $\gamma^{(i)}$, respectively) are defined by the sketch of Fig.~\ref{fig:stress_strain_sketch}. In the rest of this paper, for simplicity, the external stress $\tau^\text{ext}$ will be denoted by $\tau$.

\subsection{The SCPM model}

As said above, there are three scalar parameters of the SCPM: (i) the $\nu$ shape parameter of the Weibull distribution describing the local yield stress distribution, (ii) $\Delta \gamma^\text{pl}$ determining the local increment of the strain during the activation of a cell, and (iii) $\tau_w$ characterizing the average strength (threshold stress) of the cells. In the following section we present simulations with different parameters, and if not stated otherwise $\nu = 1.4$, $\Delta \gamma^\text{pl} = 1/4$ and $\tau_w = 1$ is used.

\subsubsection{Average stress-strain curve}
\label{sec:avg_stress_strain_scpm}

\begin{figure}[!htbp]
\begin{center}
\includegraphics[scale=0.5]{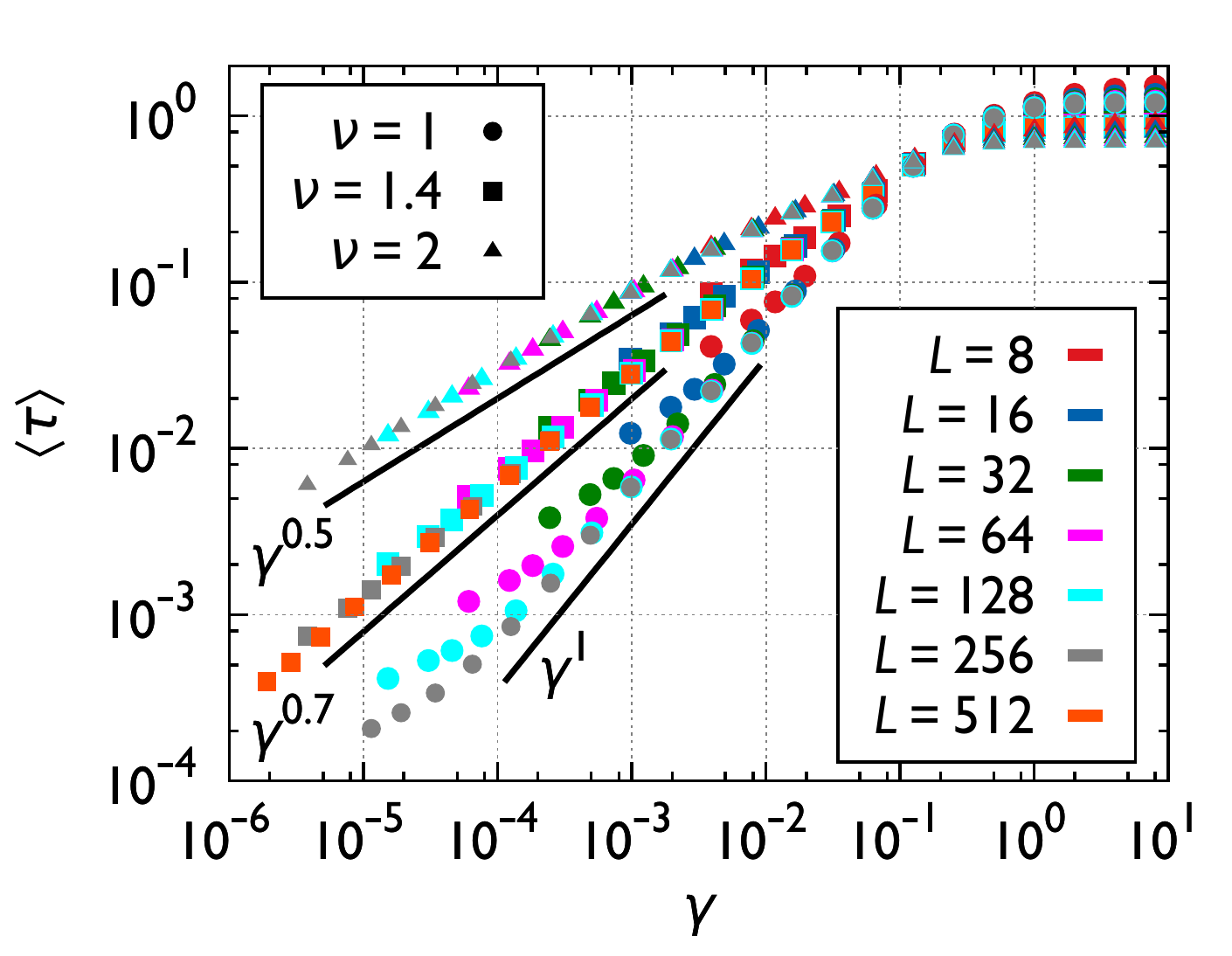}
\begin{picture}(0,0)
\put(-205,144){\textsf{(a)}}
\end{picture}
\includegraphics[scale=0.5]{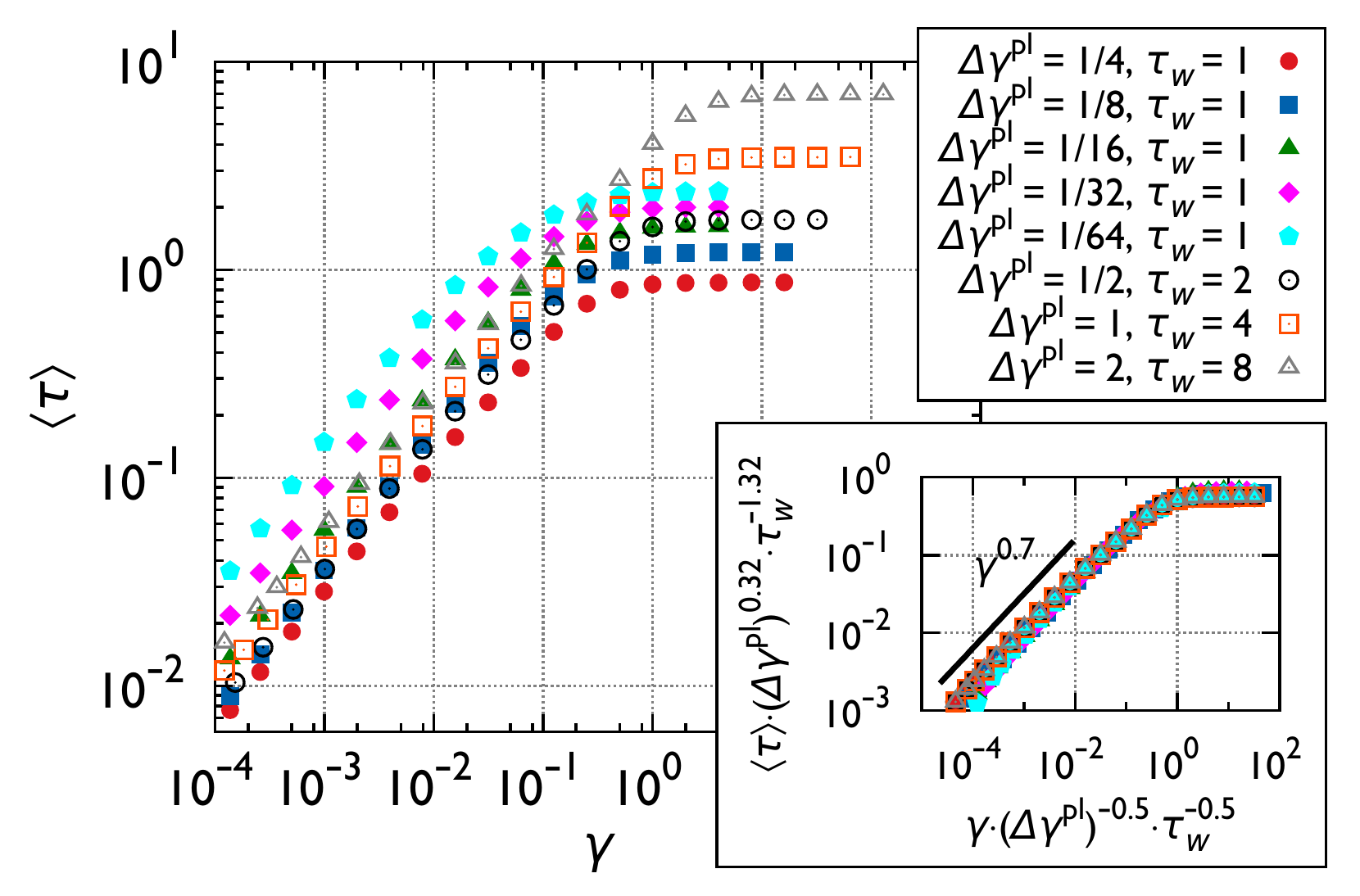}
\begin{picture}(0,0)
\put(-240,144){\textsf{(b)}}
\end{picture}
\caption{\label{fig:avg_stress_strain_SCPM} The average stress-plastic strain curves obtained by the SCPM with different choices of the parameters. In every case for small strains they follow a power-law, then saturate. (a) The effect of the shape parameter $\nu$. The power-law region is consistent with the $\langle \tau \rangle = \gamma^{1/\nu \zeta}$ relation predicted by Eq.~(\ref{eq:tau}) with $\zeta=1$. (b) The effect of $\tau_w$ and $\Delta \gamma^\text{pl}$ at $\nu=1.4$ and $L=128$. According to the scaling collapse of the inset the average stress-strain curves obey $\langle \tau \rangle = \tau_w^{1.32} (\Delta \gamma^\text{pl})^{-0.32} f(\gamma \tau_w^{-0.5} (\Delta \gamma)^{-0.5})$ with a suitable function $f$.}
\end{center}
\end{figure}
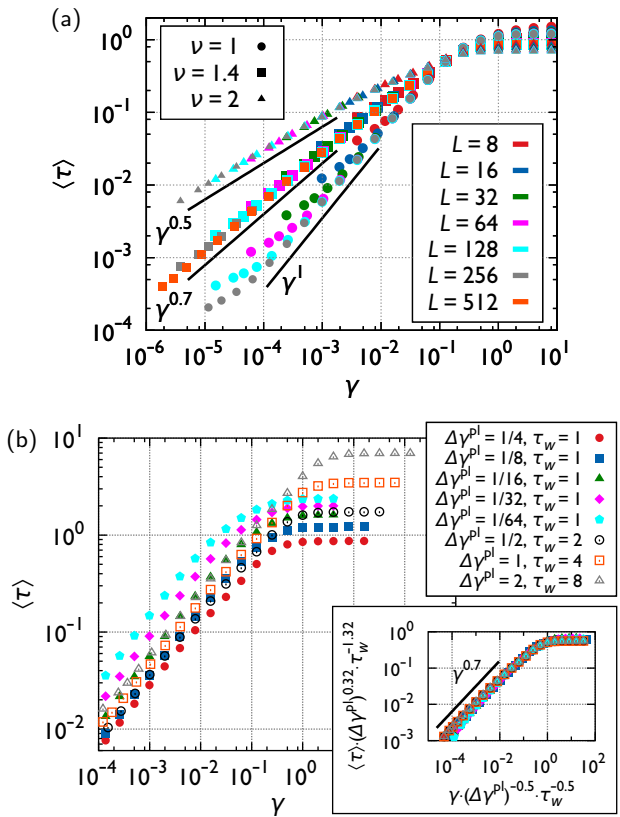

Figure \ref{fig:avg_stress_strain_SCPM}(a) plots the average stress-strain curves obtained by the SCPM simulations at different system sizes and for different values of the exponent $\nu$. The curves were obtained by the following method: for a given strain value $\gamma$ the assigned stress value $\langle \tau \rangle$ is the average of the stress values measured in individual simulations at $\gamma$. This procedure was repeated for different $\gamma$ values to obtain the whole average stress-strain curve. It is seen, that the microplastic regime is described by a power-law which follows
\begin{equation}
		\langle \tau \rangle(\gamma) = \tau_1 \gamma^\alpha
		\label{eq:avg_stress_strain}
\end{equation}
for several decades with $\tau_1$ being a constant prefactor and exponent $\alpha$ being dependent on the value of Weibull parameter $\nu$. The fitted values of $\alpha$ are summarized in Table \ref{tab:example}. It is interesting to note, that the curves do not exhibit a clear sign of size effects, they practically overlap for every $\nu$ and $L$, therefore, the $L$-dependence was neglected in Eq.~(\ref{eq:avg_stress_strain}). For plastic strains $\gamma \gtrsim 1$ the stress saturates and the system enters the continuously flowing state.

Figure \ref{fig:avg_stress_strain_SCPM}(b) shows the role of the two other parameters $\Delta \gamma^\text{pl}$ and $\tau_w$. As seen, the average stress-strain curves do depend on the specific choice, but according to the inset, a scaling collapse can be obtained if both stresses and strains are rescaled using specific powers of $\Delta \gamma^\text{pl}$ and $\tau_w$. This means that the shape of the curves is only affected by the parameter $\nu$, whereas $\Delta \gamma^\text{pl}$ and $\tau_w$ calibrates the scale of stress and strain.

\subsubsection{Fluctuation in the plastic response}

\begin{figure*}[!htbp]
\begin{center}
\hspace*{-0.5cm}
\includegraphics[scale=0.425]{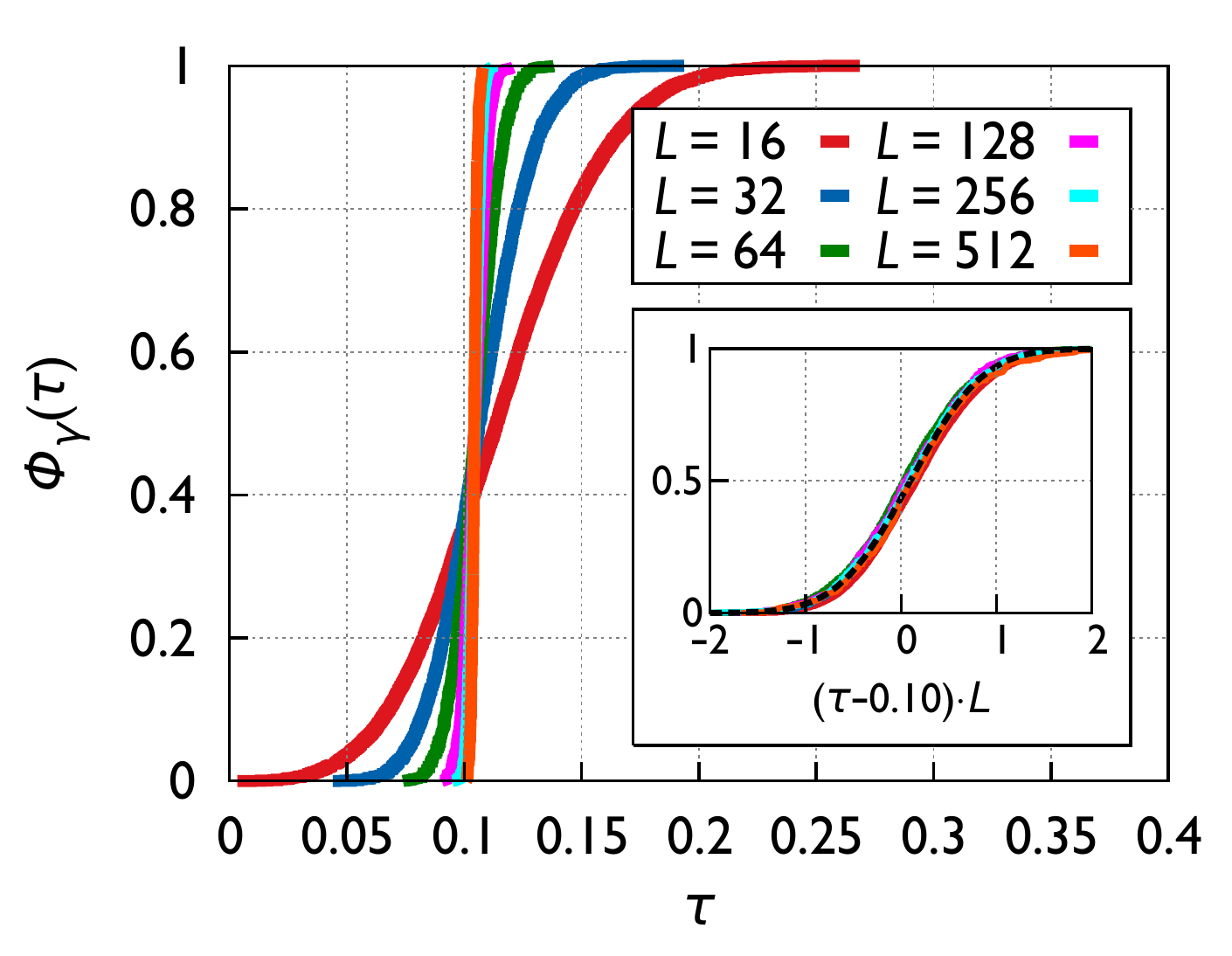}
\begin{picture}(0,0)
\put(-174,120){\textsf{(a)}}
\end{picture}
\hspace*{-0.4cm}
\includegraphics[scale=0.425]{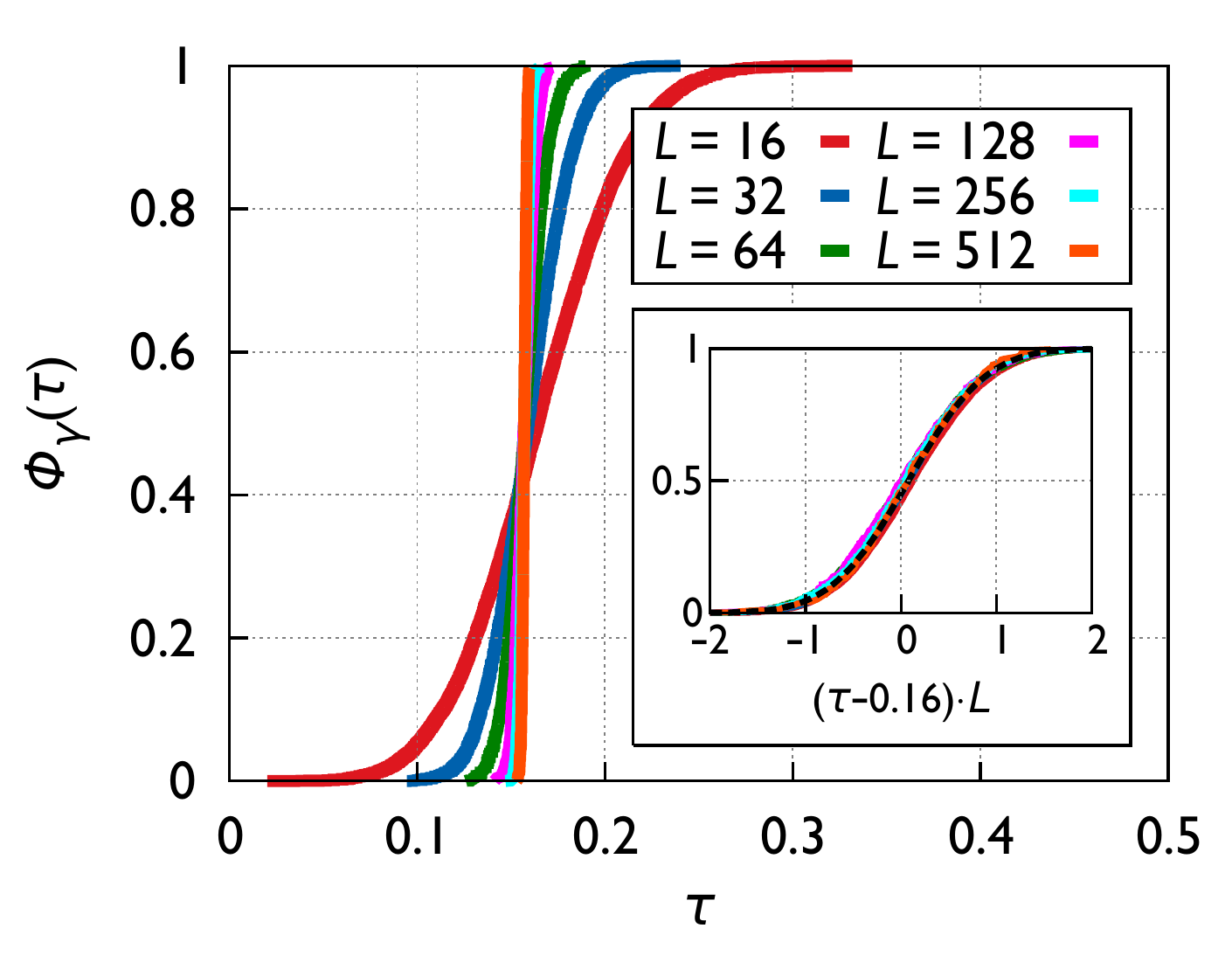}
\begin{picture}(0,0)
\put(-174,120){\textsf{(b)}}
\end{picture}
\hspace*{-0.4cm}
\includegraphics[scale=0.425]{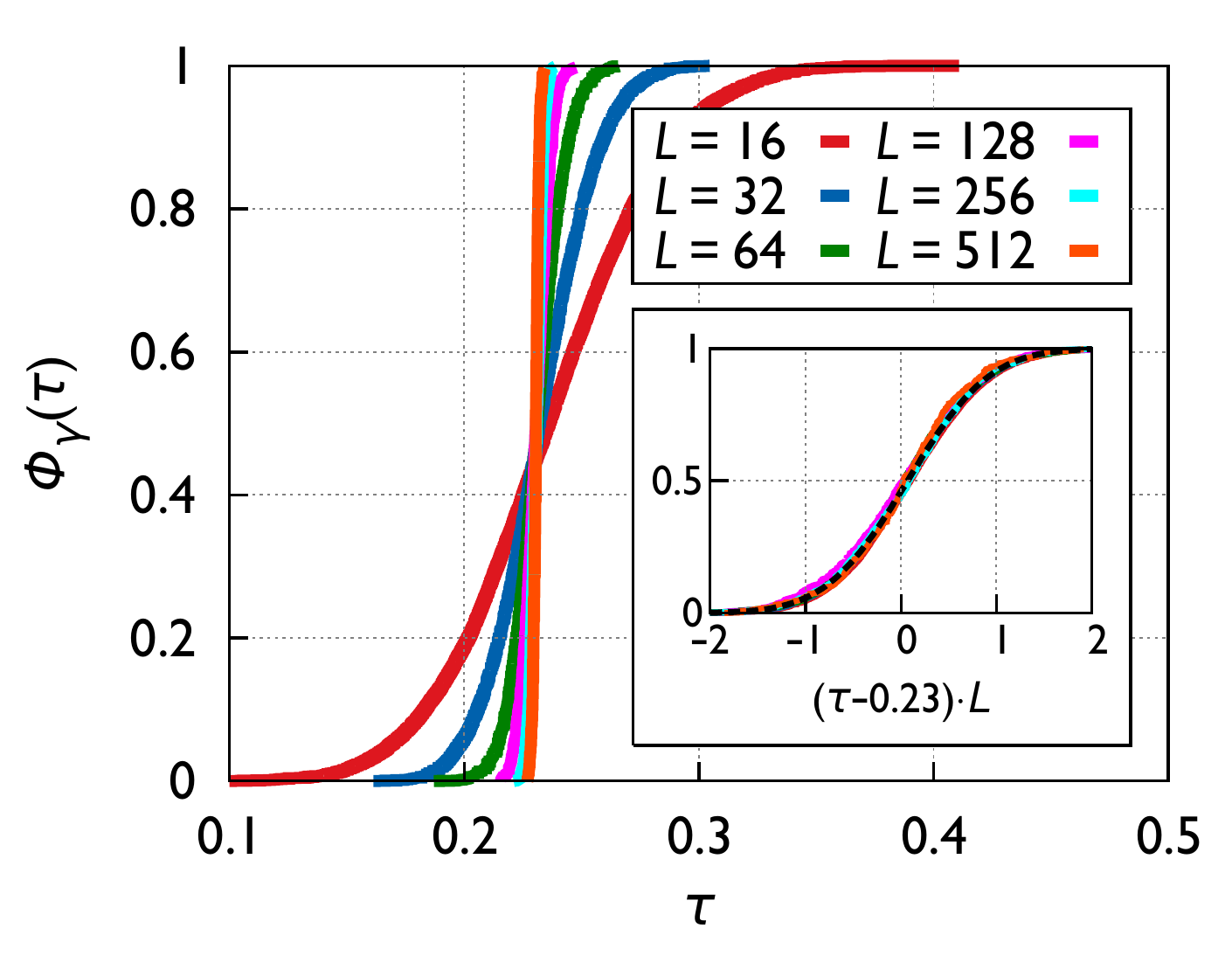}
\begin{picture}(0,0)
\put(-174,120){\textsf{(c)}}
\end{picture}
\hspace*{-0.5cm}
\caption{\label{fig:stress_distrib_scpm} Cumulative stress distribution at different deformation levels for the SCPM case. As system size increases the distributions tend to a step function, that is, stress fluctuations disappear for large samples. By multiplying the external stress with a power of the system size one can fit the curves with a normal distribution (dashed line) as can be seen in the insets. (a) $\gamma = 0.008$, (b) $\gamma = 0.016$, (c) $\gamma = 0.032$.}
\end{center}
\end{figure*}

Although the average stress-strain curve discussed above is smooth, the stress-strain curves corresponding to individual realizations are staircase-like and differ from each other. Here we investigate the cumulative distribution of stresses $\Phi_\gamma(\tau)$ measured at a given strain $\gamma$ for different realizations. The wider this distribution is, the more are the individual realizations unpredictable. Macroscopic bodies are characterized by a well-defined and smooth stress-strain curve, so for large systems one expects shrinking of this distribution. Indeed, as seen in Fig.~\ref{fig:stress_distrib_scpm} the measured $\Phi_\gamma(\tau)$ curves tend to a step function as the system size increases at every strain $\gamma$. Since there was only a negligible size effect in the average stress-strain curve, the stress-strain response of an infinite system must be equal to $\langle \tau \rangle (\gamma)$, therefore, the limiting step function must be at $\langle \tau \rangle (\gamma)$. Interestingly, the $\Phi_\gamma(\tau)$ curves seem to intersect with each other at a single point which, therefore, must correspond to $\langle \tau \rangle (\gamma)$.

According to the inset of Fig.~\ref{fig:stress_distrib_scpm} the curves can be collapsed by rescaling the stresses by the system size around $\langle \tau \rangle (\gamma)$. In addition, the curves can be fit very well with a normal distribution, that is,
\begin{equation}
	\Phi_\gamma(\tau) = \frac12 \left[ 1 + \text{erf} \left( \frac{\tau-\langle \tau \rangle (\gamma)}{c L^{-\beta}} \right) \right],
	\label{eq:stress_fluct}
\end{equation}
where $\langle \tau \rangle (\gamma)$ is the average stress-strain curve of Eq.~(\ref{eq:avg_stress_strain}), $\beta = 1 \pm 0.05$ is the exponent characterizing the system size dependence of the stress fluctuations, and $c$ is an appropriate constant.

We note that to fit the same distribution for 2D and 3D DDD as well as micro-pillar compression data a shifted Weibull distribution was used previously with shape parameter $\sim$3.5.\cite{Ispanovity2013} Since a Weibull with this shape parameter is practically indistinguishable from a normal distribution, we used here the latter, as it contains only two fitting parameters. It is also noted that the theory to be proposed in Sec.~\ref{sec:model} predicts the normal distribution of Eq.~(\ref{eq:stress_fluct}).

\subsubsection{The stress sequence}

\begin{figure}[!htbp]
\begin{center}
\includegraphics[scale=0.5]{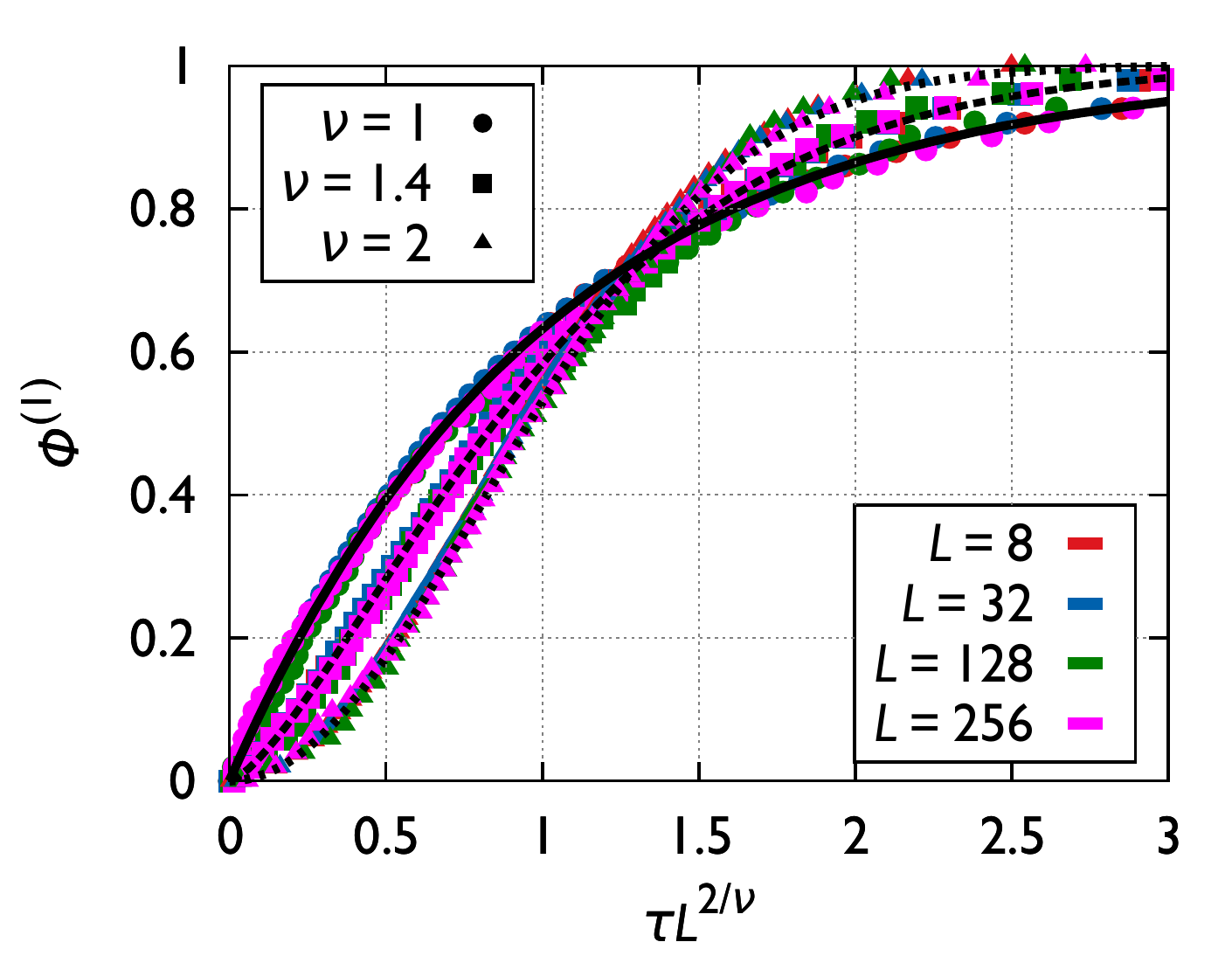}
\begin{picture}(0,0)
\put(-205,144){\textsf{(a)}}
\end{picture}
\includegraphics[scale=0.5]{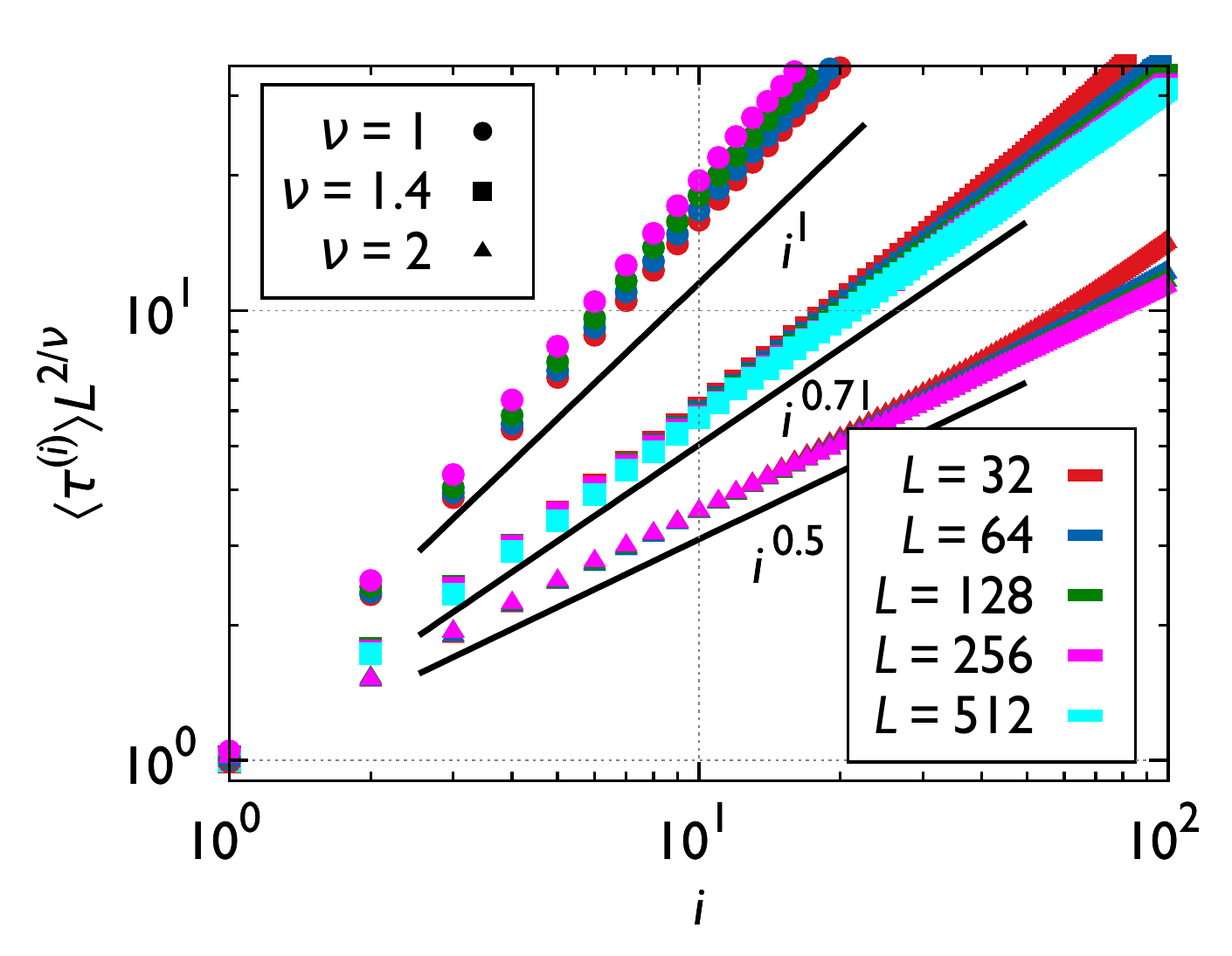}
\begin{picture}(0,0)
\put(-205,144){\textsf{(b)}} 
\end{picture}
\includegraphics[scale=0.5]{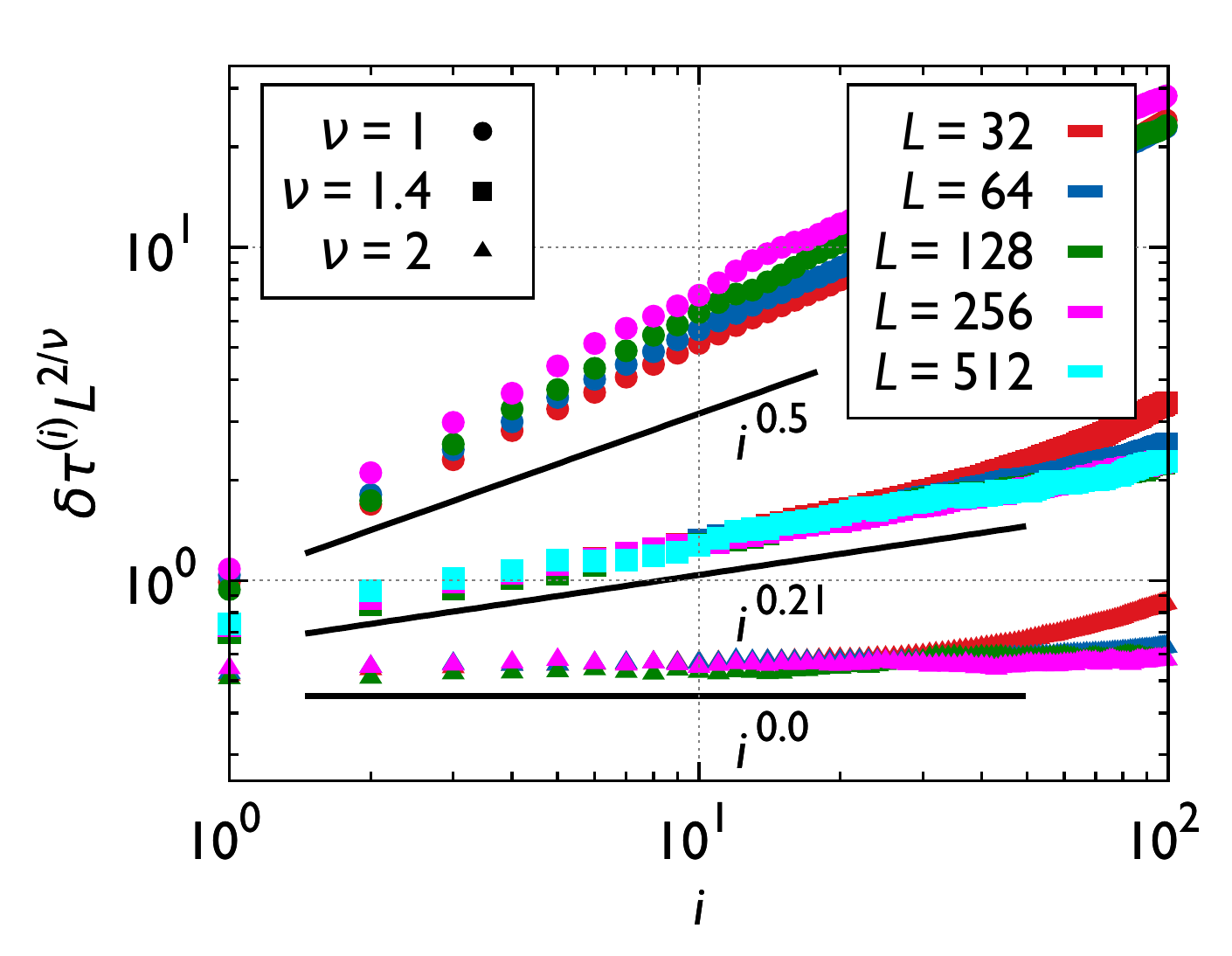}
\begin{picture}(0,0)
\put(-205,144){\textsf{(c)}} 
\end{picture}
\caption{\label{fig:tau_i_scpm} (a) The cumulative distribution $\Phi^{(1)}$ of the first activation stress $\tau^{(1)}$ for SCPM simulations with $\nu=1, 1.4, 2$ and different system sizes. The scaling collapse for different system sizes is obtained by rescaling the stress by $L^{2/\nu}$. The corresponding Weibull distributions of Eq.~(\ref{eqn:weibull}) are also plotted (solid, dashed, and dotted lines). (b) The average stress sequence $\langle \tau^{(i)} \rangle$. The curves are proportional with $i^{1/\nu}$ and the ones corresponding to different system sizes collapse if stresses are rescaled by $L^{2/\nu}$, in accordance with Eq.~(\ref{eq:tau_average_scpm}). (c) STD of the stress sequence $\delta \tau^{(i)}$. The curves are consistent with Eq.~(\ref{eq:tau_std_scpm}).
}
\end{center}
\end{figure}

The two following subsections aim at studying the statistics of the stress and strain sequences $\tau^{(i)}$ and $\gamma^{(i)}$, because they will play a central role in the simple plasticity model described in Sec.~\ref{sec:model}. First, the distribution $\Phi^{(1)}$ of the stress where the first event takes place $\tau^{(1)}$ is considered. In the SCPM model the plastic strain field is initially zero, therefore, the local stress is everywhere equal to the applied stress until the occurrence of the first event. Consequently, the distribution of the stress where the first plastic event sets on $\Phi^{(1)}(\tau^{(1)})$ must be described by a Weibull distribution with shape parameter $\nu$ and scale parameter proportional with $L^{-2/\nu}$ (see Sec.~\ref{sec:stress_seq} for details). Indeed, according to Fig.~\ref{fig:tau_i_scpm}(a) $\Phi^{(1)}$ is perfectly fit by the corresponding Weibull distribution, and the distributions overlap if stress values are rescaled by $L^{2/\nu}$.

Figures \ref{fig:tau_i_scpm}(b) and \ref{fig:tau_i_scpm}(c) plot the average stress sequence $\langle \tau^{(i)} \rangle$ and its standard deviation (STD) $\delta \tau^{(i)}$, respectively. The curves corresponding to a given $\nu$ parameter and for small $i$ values (that is, when $\tau^{(i)} \lesssim 0.1$) overlap if the stresses are rescaled by $L^{2/\nu}$ and are well described by the power-laws
\begin{align}
	\langle \tau^{(i)} \rangle = \tau_0 \left( \frac{i}{L^\eta} \right)^{1/\nu}, 
	\label{eq:tau_average_scpm}\\
	\delta \tau^{(i)} = \frac{\tau_0}{i^{1/2}} \left( \frac{i}{L^\eta} \right)^{1/\nu},
	\label{eq:tau_std_scpm}
\end{align}
with $\eta=2.0\pm 0.05$ yielded by visual inspection.

\subsubsection{Strain sequence}

\begin{figure}[!htbp]
\begin{center}
\includegraphics[scale=0.5]{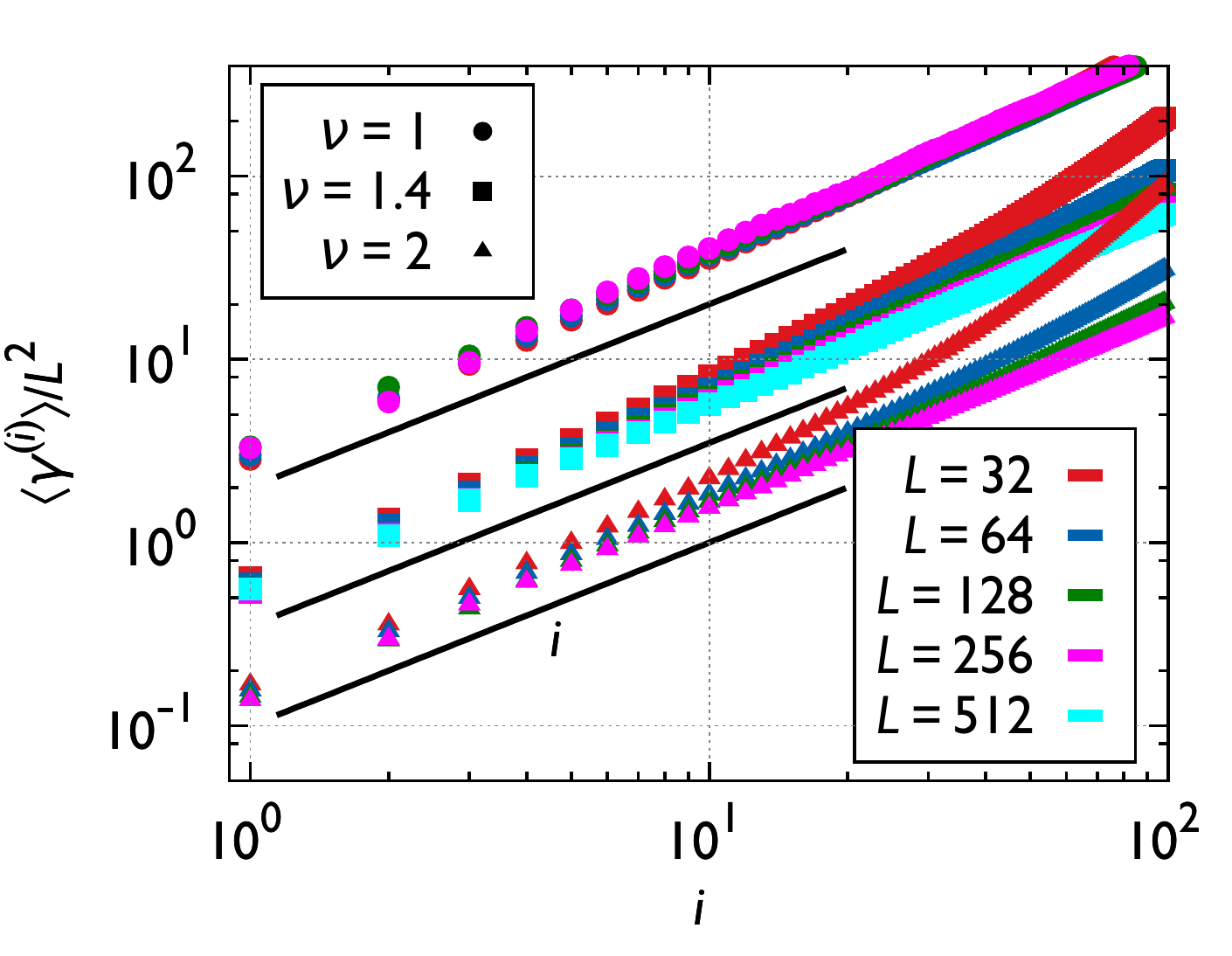}
\begin{picture}(0,0)
 \put(-205,144){\textsf{(a)}}
\end{picture}
\includegraphics[scale=0.5]{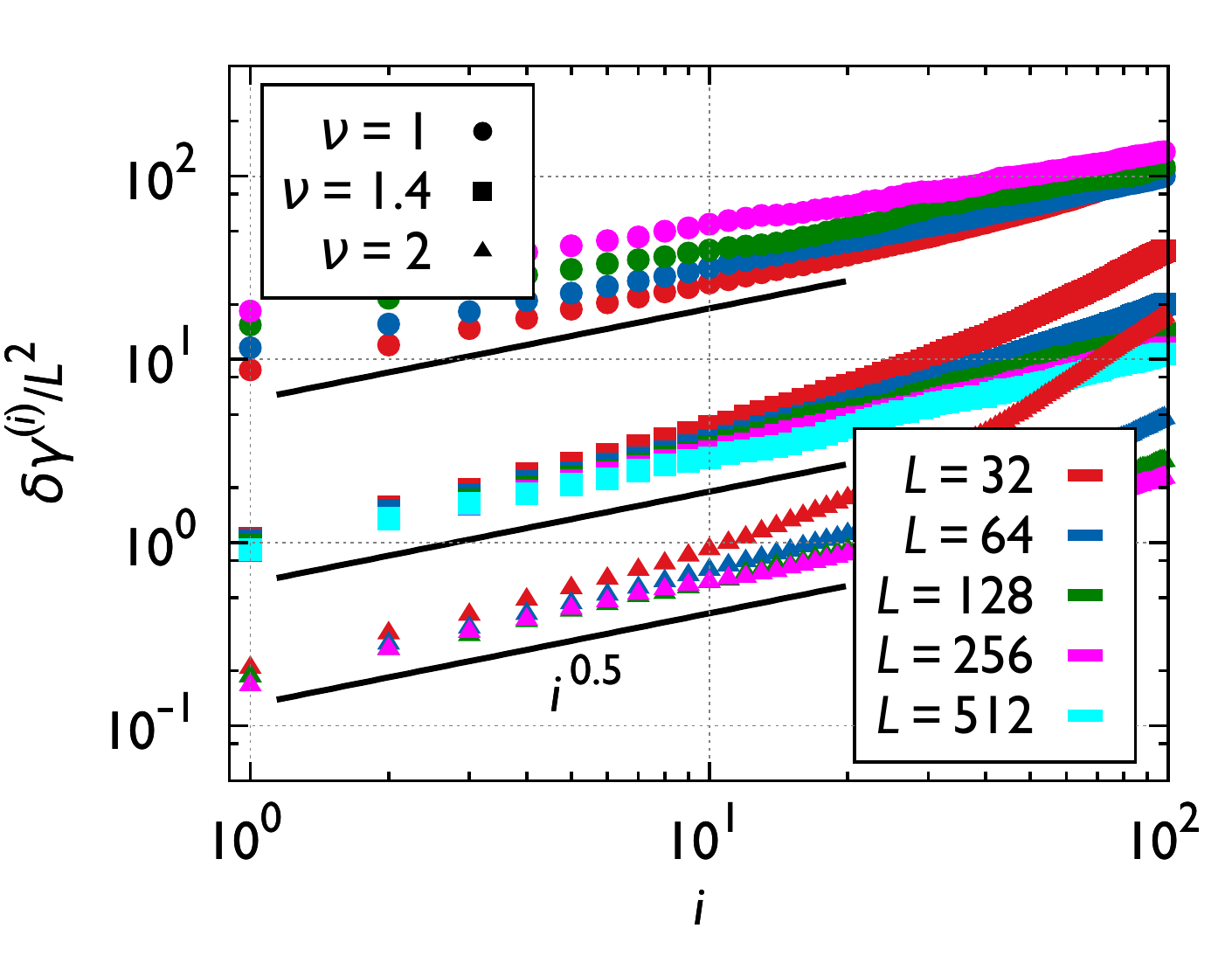}
\begin{picture}(0,0)
 \put(-205,144){\textsf{(b)}}
\end{picture}
\caption{\label{fig:gamma_i_scpm} The average [panel (a)] and STD [panel (b)] of the strain $\gamma^{(i)}$ measured at the $i$th strain burst for different system sizes and $\nu$ values. The curves follow Eqs.~(\ref{eq:strain_sequence_average}) and (\ref{eq:strain_sequence_std}) with $\zeta=1$ and $\xi=2$. 
}
\end{center}
\end{figure}

The average and the STD of the strain sequence $\gamma^{(i)}$ is seen in Fig.~\ref{fig:gamma_i_scpm}. It is clear that both $\langle \gamma^{(i)} \rangle$ and $\delta \gamma^{(i)}$ follow a power-law for small $i$ values:
\begin{align}
	\langle \gamma^{(i)} \rangle = s_0 \frac{i^\zeta}{L^\xi},
	\label{eq:strain_sequence_average}  \\
	\delta \gamma^{(i)} = s_1 \frac{i^{\zeta-1/2}}{L^\xi},
	\label{eq:strain_sequence_std}
\end{align}
with $\zeta=1.0\pm 0.05$ and $\xi = 2.0 \pm 0.1$. It is important to note, that none of the exponents $\zeta$ and $\nu$ are sensitive to the choice of the exponent $\nu$, but the actual level of $\gamma^{(i)}$ and its scatter (and, thus, the values $s_0$ and $s_1$) are significantly larger for smaller $\nu$ values, which means that individual avalanches become much larger in this case. 

\subsection{DDD models}

\subsubsection{Average stress-strain curve}

\begin{figure}[!htbp]
\begin{center}
\includegraphics[scale=0.5]{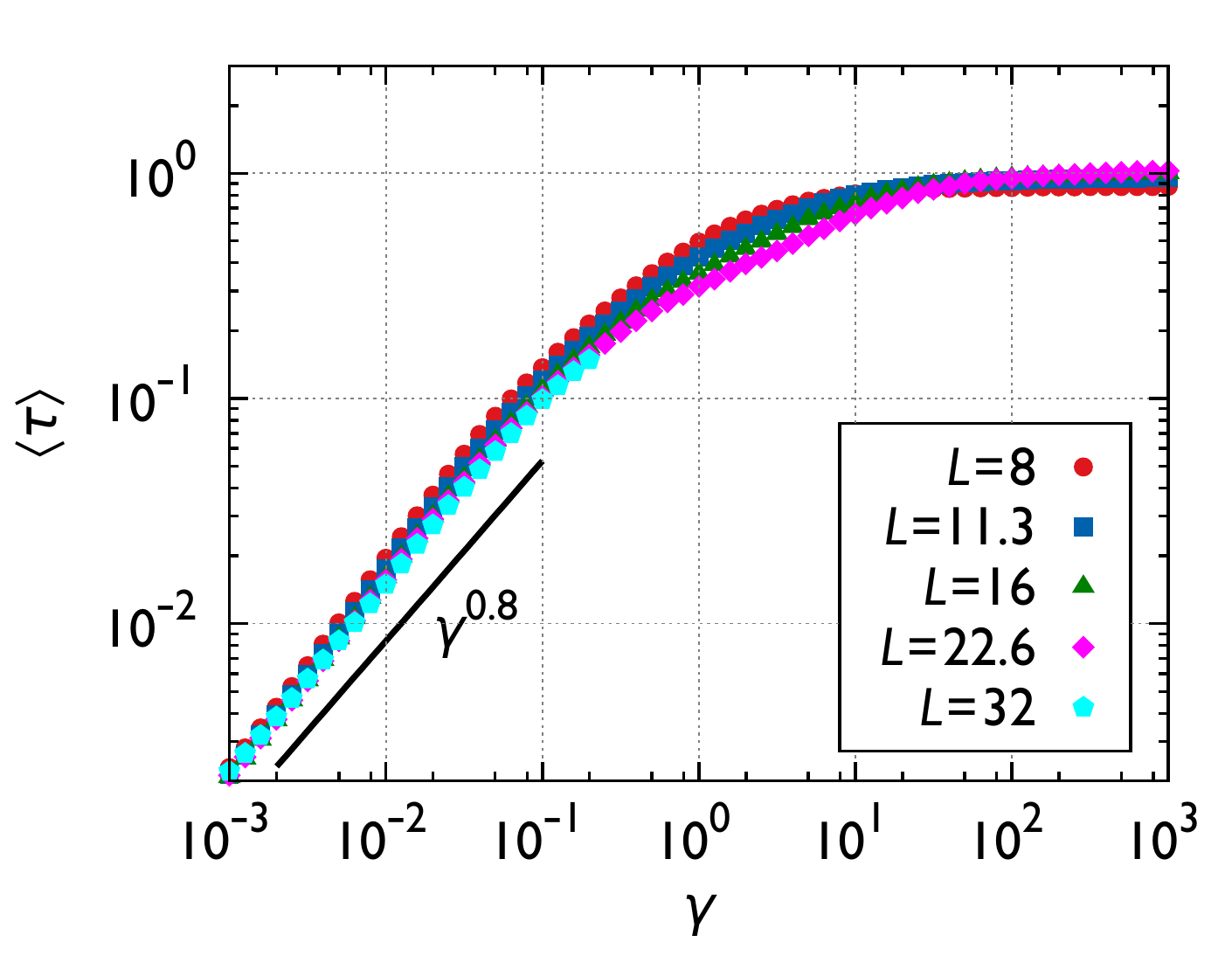}
\begin{picture}(0,0)
 \put(-205,144){\textsf{(a)}}
\end{picture}
\includegraphics[scale=0.5]{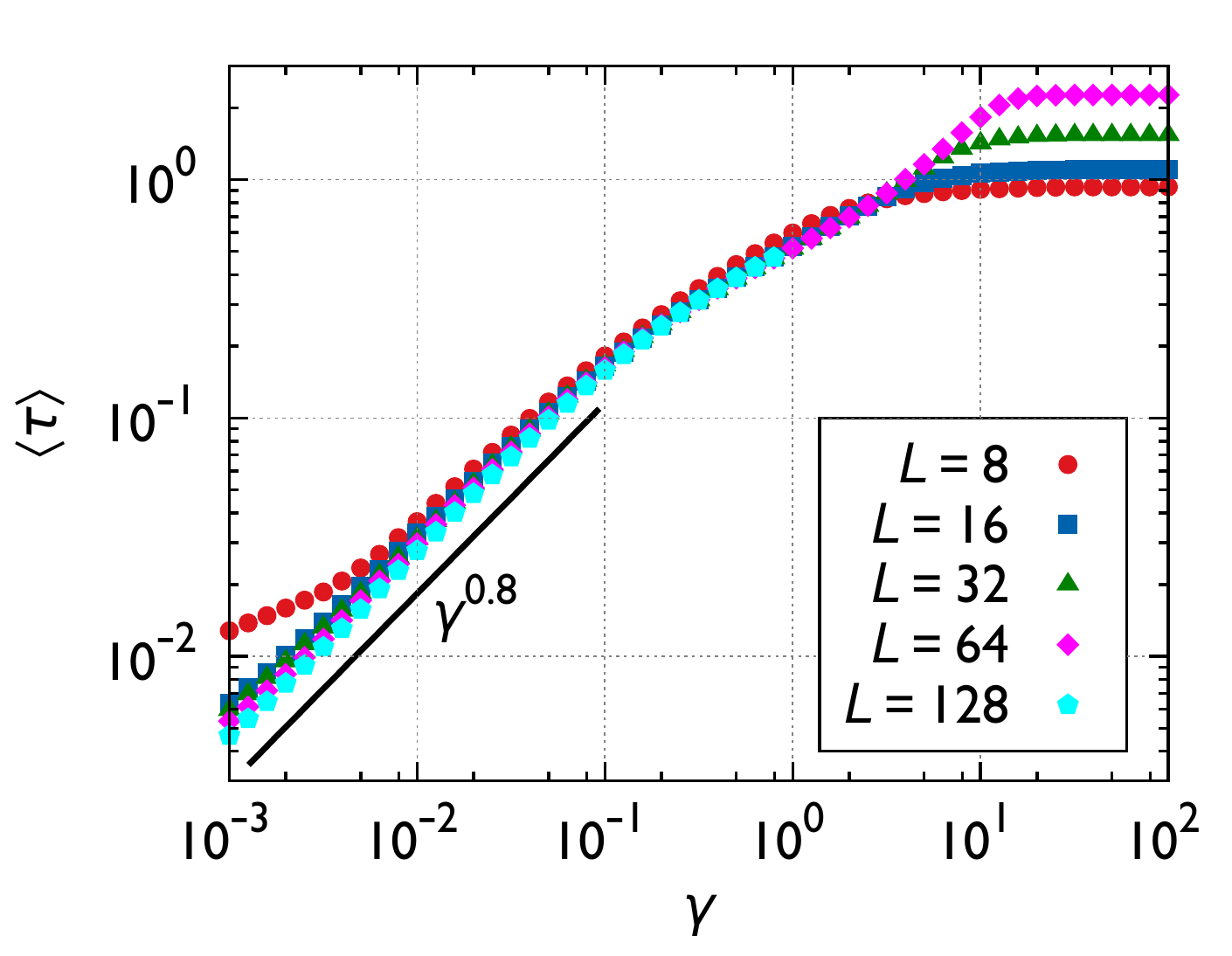}
\begin{picture}(0,0)
 \put(-205,144){\textsf{(b)}}
\end{picture}
\caption{\label{fig:avg_stress_strain_DDD} The average stress-strain curves of the DDD simulations. They follow a power-law until $\gamma \approx 0.1$, then saturate. The panels correspond to (a) TCDDD, (b) CADDD simulations.}
\end{center}
\end{figure}

The average plastic response of the specimens was calculated in the same manner as for SCPM described in Sec.~\ref{sec:avg_stress_strain_scpm}. According to Fig.~\ref{fig:avg_stress_strain_DDD} the average stress-strain curves show similar features to those obtained by the SCPM: (i) the microplastic regime is characterized by a power-law with an exponent $\alpha = 0.8 \pm 0.05$ [see Eq.~(\ref{eq:avg_stress_strain})] and only a weak sign of size effects is seen, (ii) this regime breaks down at $\tau \approx 0.1$, and (iii) the stress-strain curves saturate for large ($\gamma \gtrsim 1$) strains.

The only significant difference in the average stress-strain curves between the different DDD simulations appears at large ($\gamma \gtrsim 1$) strains. Here the spatial discretization of the CA model leads to an increasing stress for large systems. In this part of the stress-strain curve mechanisms not included in these simple models (like dislocation creation or cross-slip) may also play an important role, so in this paper we focus on small to medium strains, where the two models yield quite similar behavior.

\subsubsection{Fluctuation in the plastic response}

\begin{figure*}[!htbp]
\begin{center}
\hspace*{-0.4cm}
\includegraphics[scale=0.425]{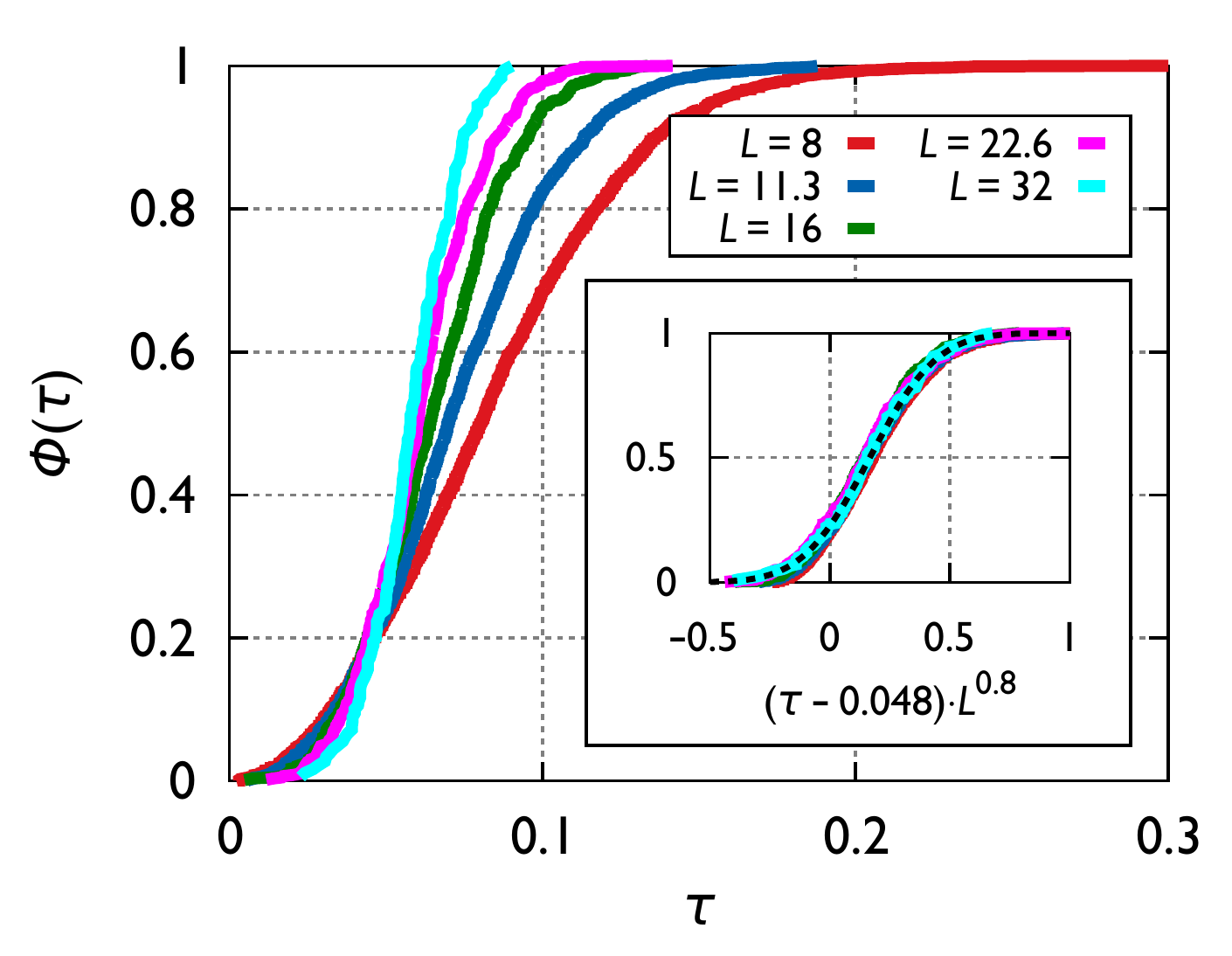}
\begin{picture}(0,0)
\put(-174,120){\textsf{(a)}}
\end{picture}
\hspace*{-0.4cm}
\includegraphics[scale=0.425]{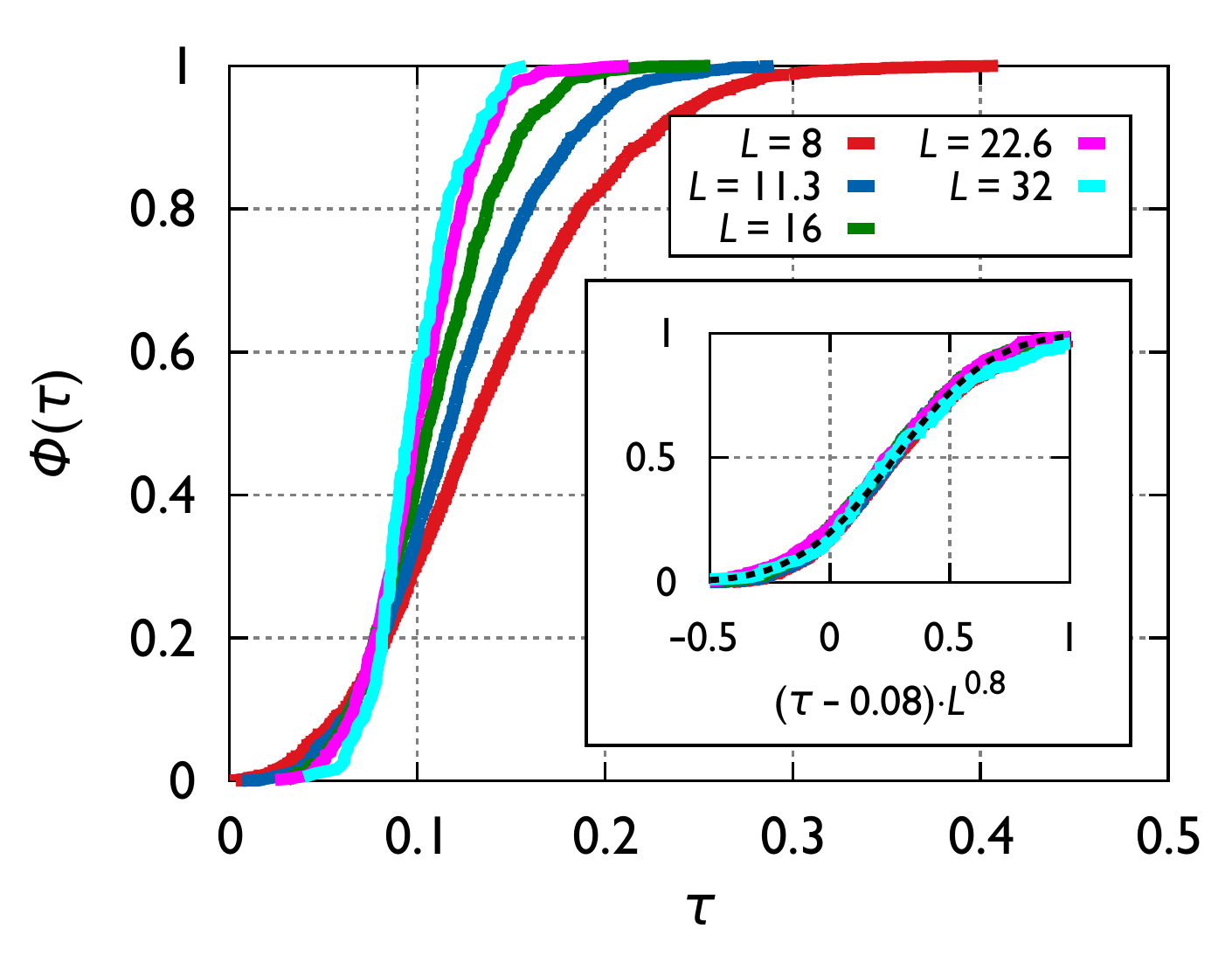}
\begin{picture}(0,0)
\put(-174,120){\textsf{(b)}}
\end{picture}
\hspace*{-0.4cm}
\includegraphics[scale=0.425]{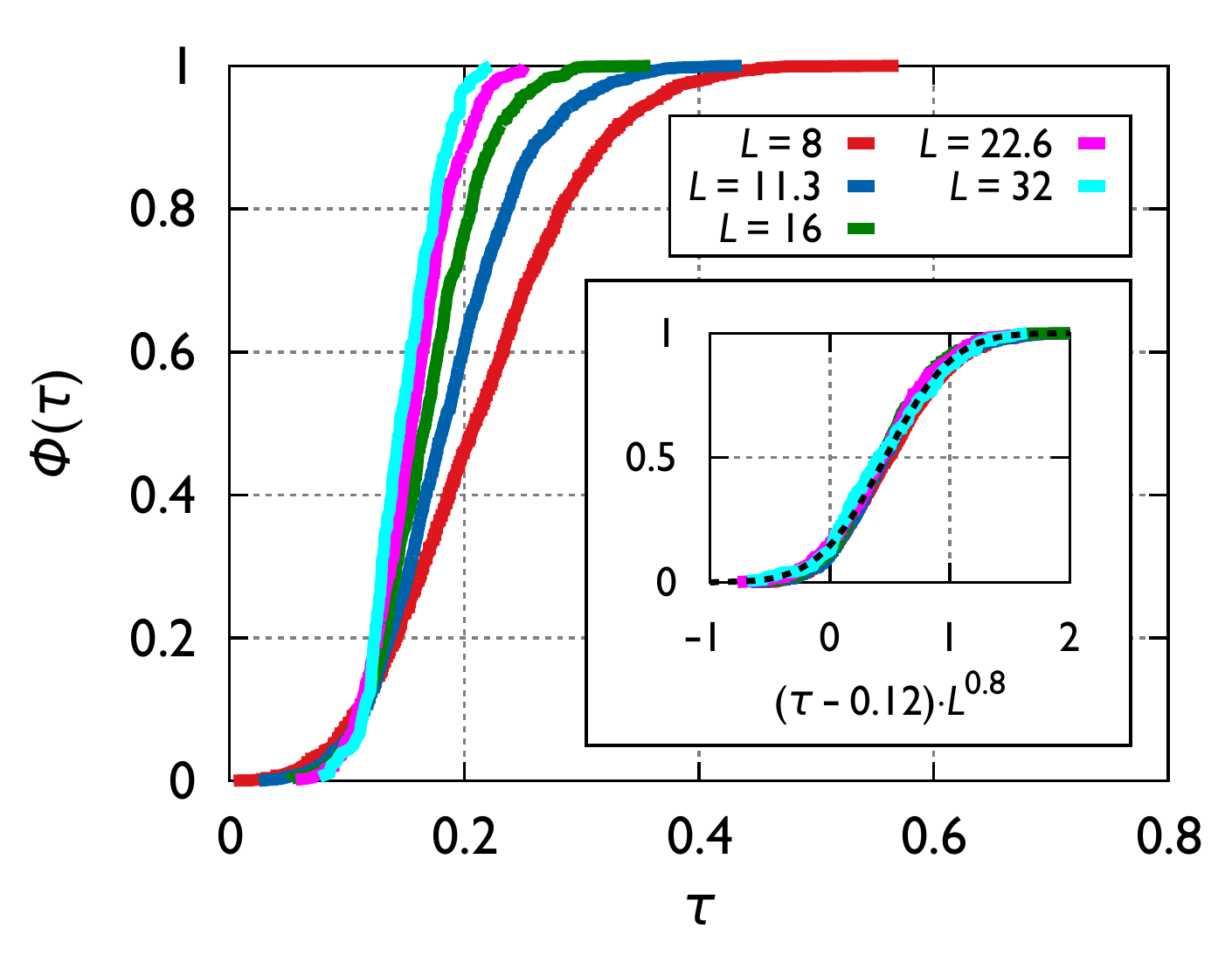}
\begin{picture}(0,0)
\put(-174,120){\textsf{(c)}}
\end{picture}
\hspace*{-0.4cm}
\includegraphics[scale=0.425]{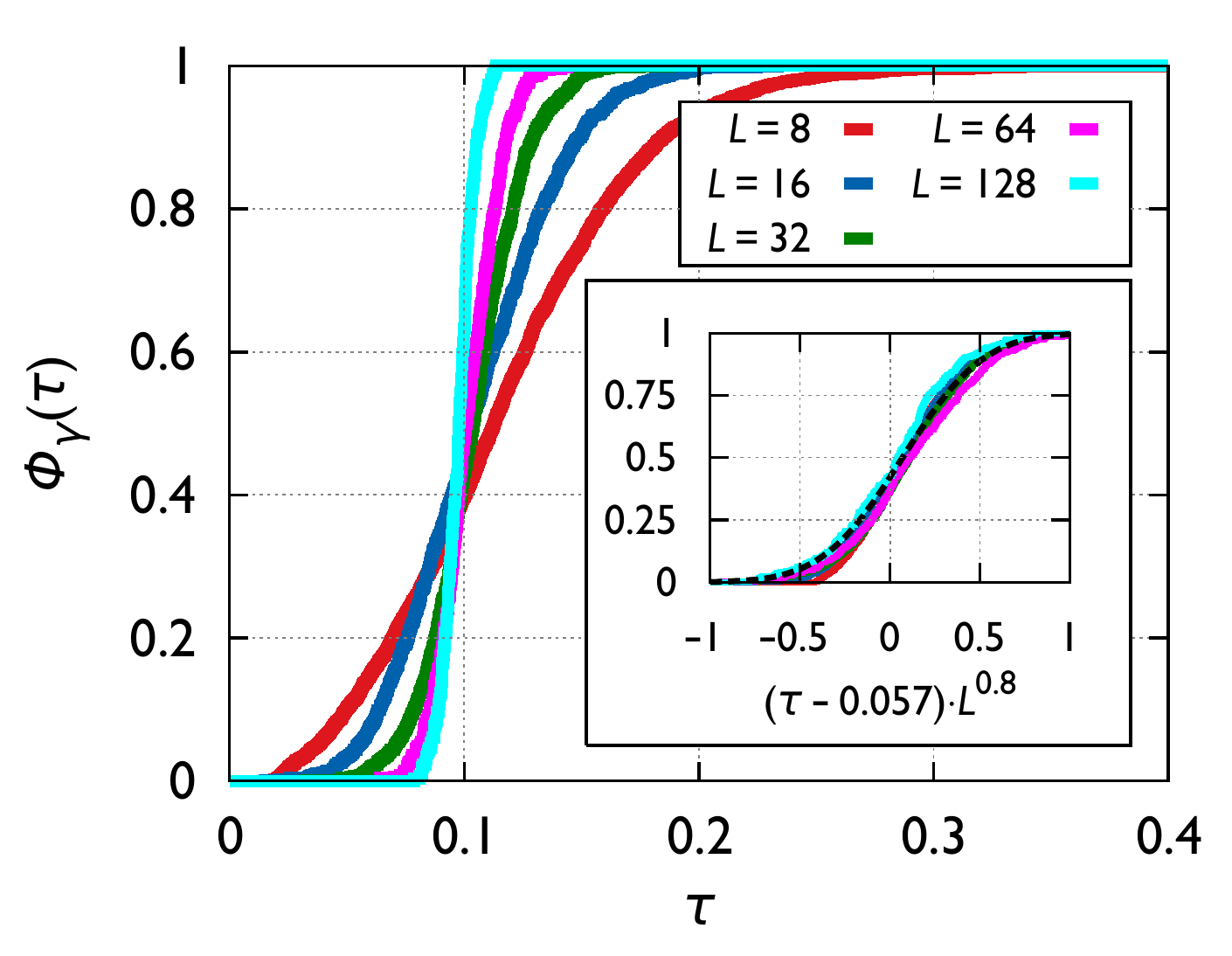}
\begin{picture}(0,0)
\put(-174,120){\textsf{(d)}}
\end{picture}
\hspace*{-0.4cm}
\includegraphics[scale=0.425]{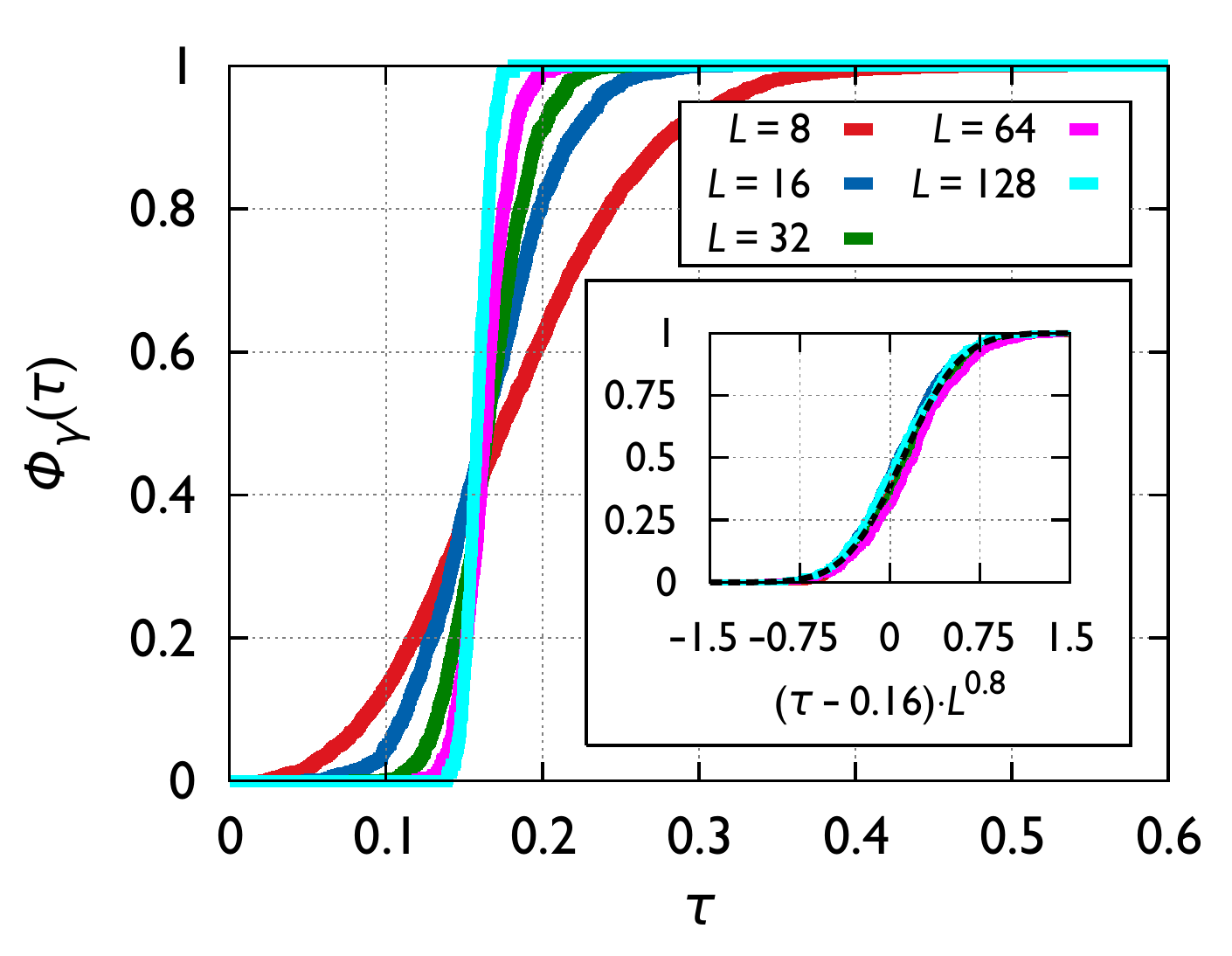}
\begin{picture}(0,0)
\put(-174,120){\textsf{(e)}}
\end{picture}
\hspace*{-0.4cm}
\includegraphics[scale=0.425]{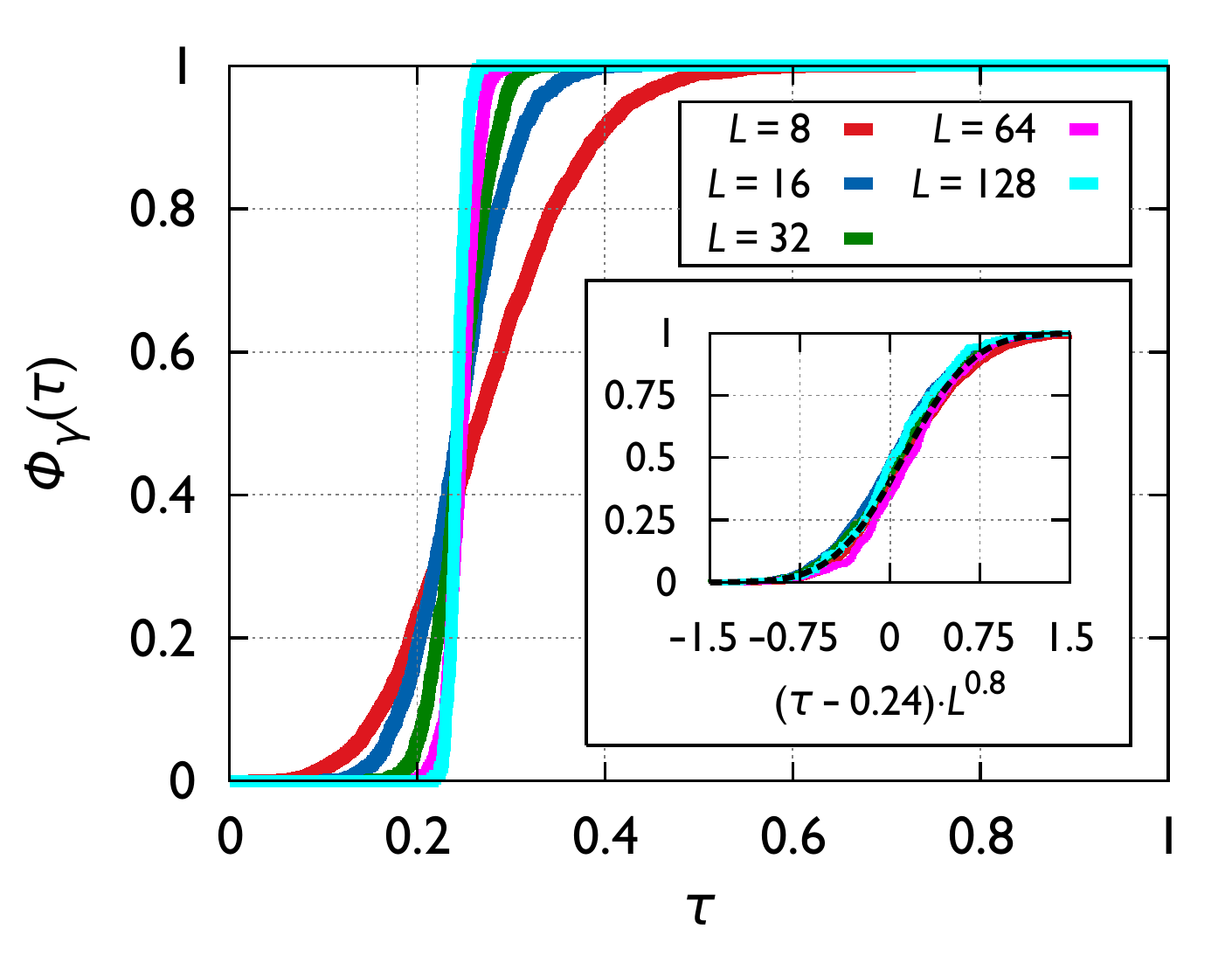}
\begin{picture}(0,0)
\put(-174,120){\textsf{(f)}}
\end{picture}
\caption{\label{fig:stress_distrib_ddd} Cumulative stress distributions $\Phi_\gamma$ at different deformation levels $\gamma$ for the DDD cases. As in Fig.~\ref{fig:stress_distrib_scpm} scaling collapse can be obtained by multiplying the external stress with a power of the system size. The so collapsed curves can be fit by an appropriate normal distribution (dashed lines). Panels (a)-(c) and (d)-(f) correspond to TCDDD and CADDD, respectively. (a),(d): $\gamma = 0.05$, (b),(e): $\gamma=0.1$, (c),(f): $\gamma=0.2$.}
\end{center}
\end{figure*}

The cumulative distribution $\Phi_\gamma$ of the stresses measured for different realizations at a given strain $\gamma$ are plotted in Fig.~\ref{fig:stress_distrib_ddd}. Like in the case of SCPM, they (i) tend to a step function for large systems, (ii) the curves for different system sizes intersect in a single point, (iii) by scaling the stresses with a power of the system size scaling collapse can be obtained, and (iv) the curves follow a normal distribution. This means that Eq.~(\ref{eq:stress_fluct}) is valid now with $\beta = 0.8 \pm 0.05$, being somewhat smaller than the value obtained for SCPM.

\subsubsection{The stress sequence}

\begin{figure}[!htbp]
\begin{center}
\includegraphics[scale=0.5, angle=0]{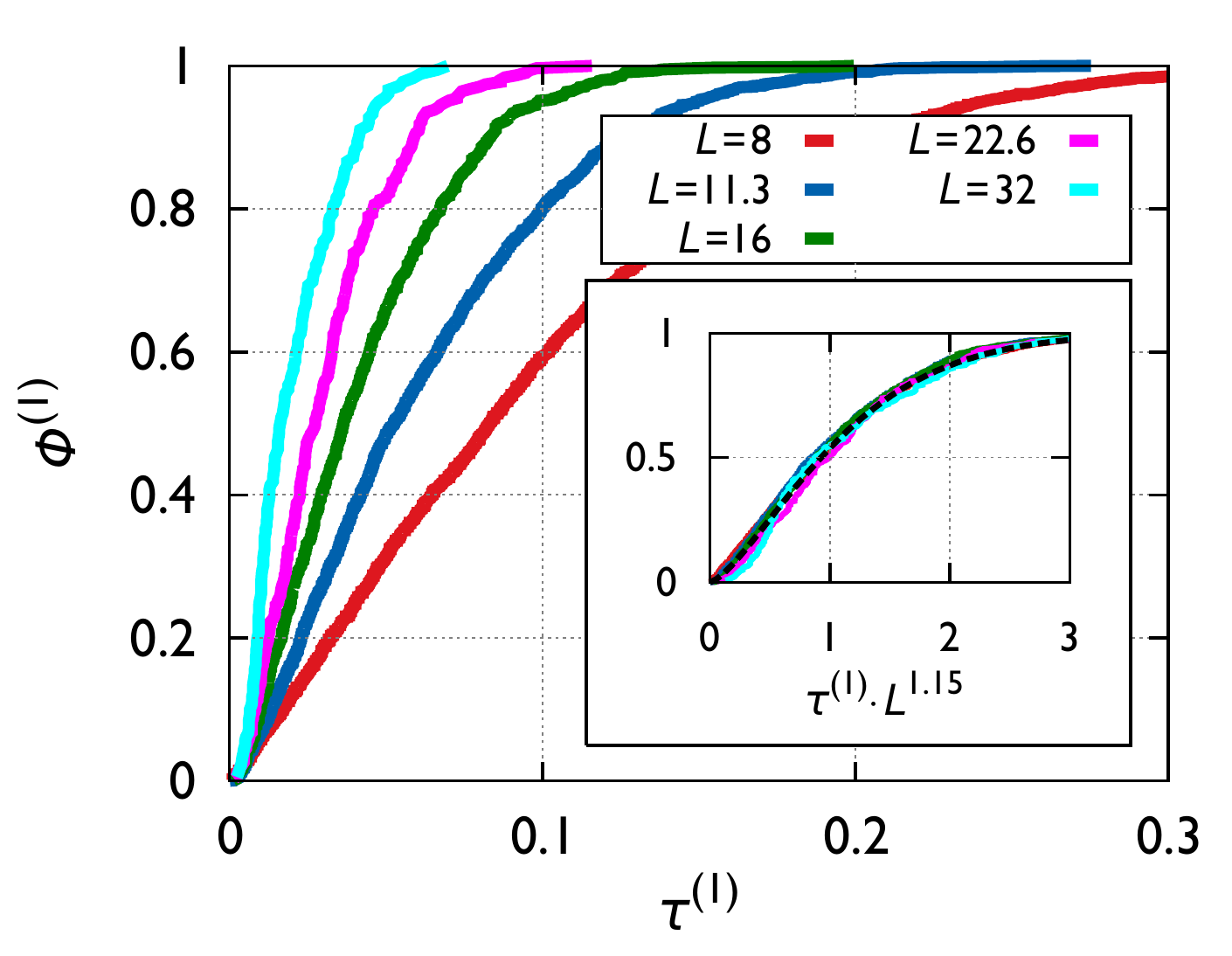}
\begin{picture}(0,0)
 \put(-205,144){\textsf{(a)}}
\end{picture}
\includegraphics[scale=0.5, angle=0]{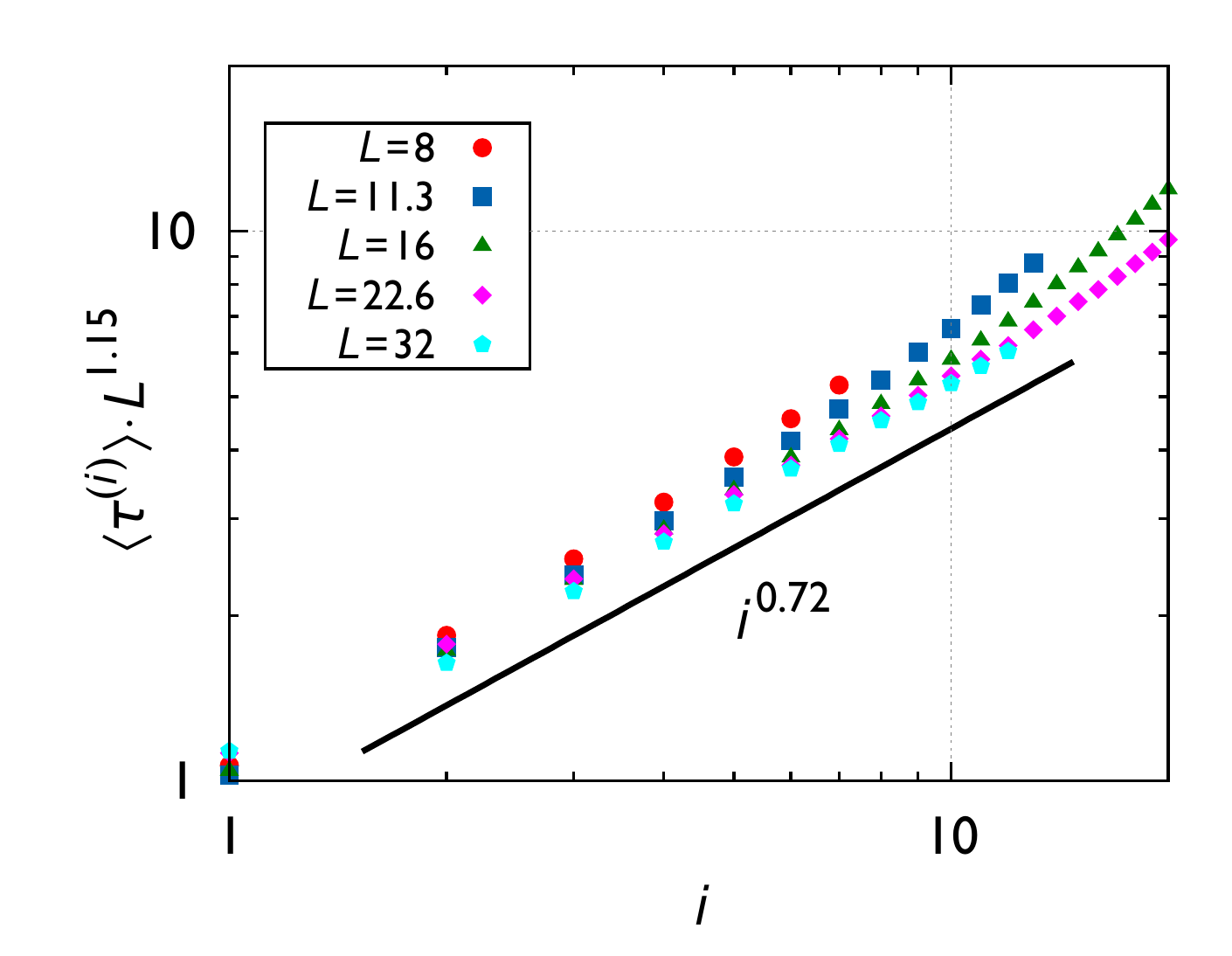}
\begin{picture}(0,0)
 \put(-205,144){\textsf{(b)}}
\end{picture}
\includegraphics[scale=0.5, angle=0]{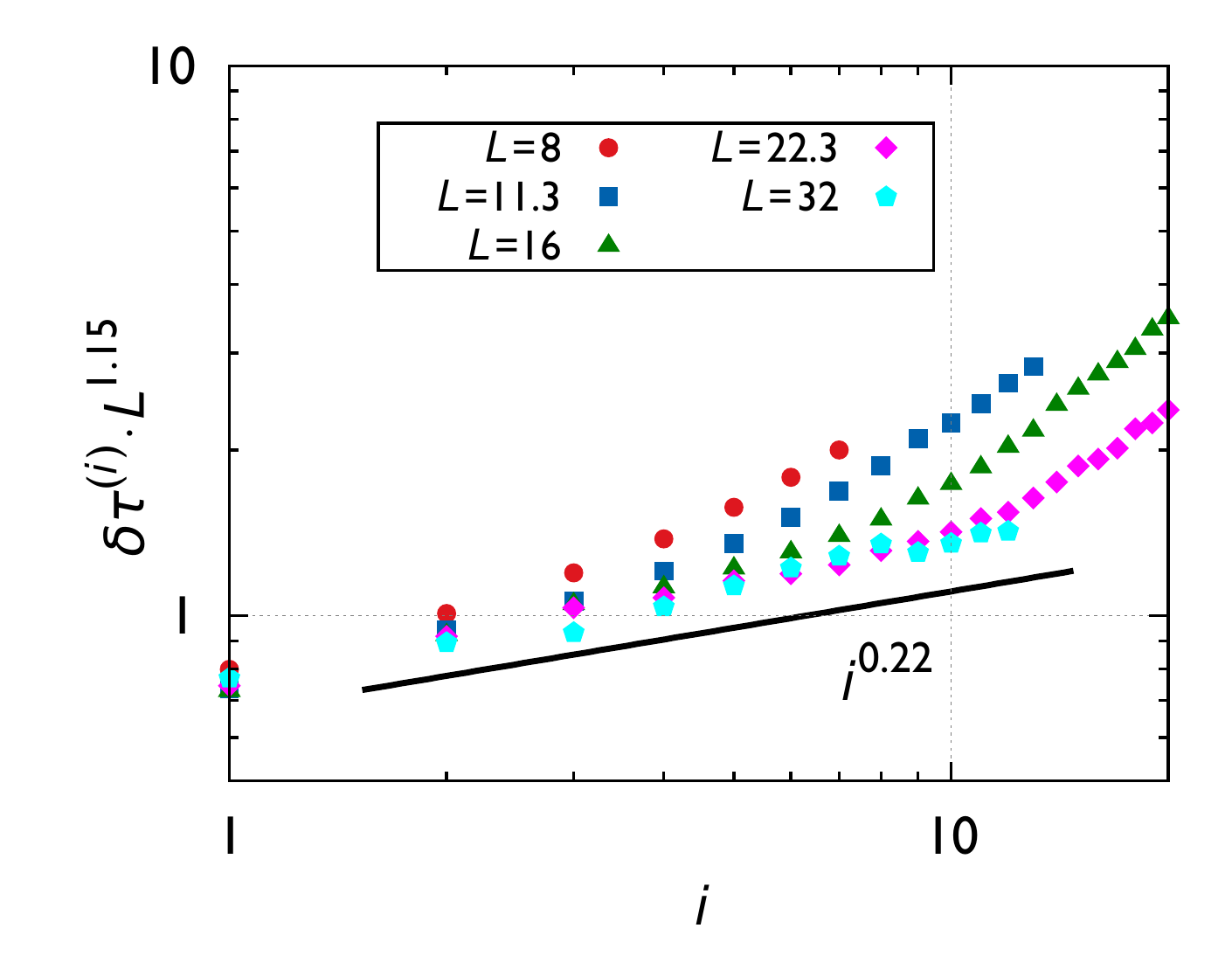}
\begin{picture}(0,0)
 \put(-205,144){\textsf{(c)}}
\end{picture}
\caption{\label{fig:tau_i_tcddd} Statistics of the stress sequence $\tau^{(i)}$ for TCDDD simulations of different system sizes.\\
(a) The cumulative distribution $\Phi^{(1)}$ of the first activation stress $\tau^{(1)}$. Inset: data collapse is obtained by plotting $\Phi^{(1)}$ as a function of $\tau L^{1.15}$. The curves can be fitted with a Weibull-distribution with a shape parameter $\nu \approx 1.4 \pm 0.05$.\\
(b),(c) The average and STD of the external stress $\tau^{(i)}$ at the \textit{i}th avalanche for different system sizes. The data are consistent with Eqs.~(\ref{eq:tau_average_scpm}) and (\ref{eq:tau_std_scpm}) (solid lines) if $i\gtrsim3$ and $\langle \tau^{(i)}\rangle \lesssim 0.2$ and with $\nu = 1.4 \pm 0.05$ and $\eta = 1.6 \pm 0.1$.}
\end{center}
\end{figure}

The following two subsections investigate the statistics of stress and strain sequences introduced above. As said in Sec.~\ref{sec:caddd}, these sequences cannot be unambiguously defined for CADDD, we, therefore, constrain ourselves to the TCDDD simulations. First, like for the SCPM, the cumulative distribution $\Phi^{(1)}$ of $\tau^{(1)}$, i.e.,\ the stress where the first plastic event sets on, is calculated. According to Fig.~\ref{fig:tau_i_tcddd}(a) $\Phi^{(1)}$ can be fit perfectly by a Weibull distribution with shape parameter $\nu$ that can be collapsed for different system sizes when rescaled by $L^{\eta/\nu}$, with parameters
\begin{align}
\nu = 1.4 \pm 0.05,\\
\eta = 1.6 \pm 0.1
\end{align}

Similarly to the SCPM case, the average $\langle \tau^{(i)} \rangle$ and STD $\delta \tau^{(i)}$ follow Eqs.~(\ref{eq:tau_average_scpm}) and (\ref{eq:tau_std_scpm}), respectively, with the same exponents $\nu$ and $\eta$ obtained from $\Phi^{(1)}$ above.

\subsubsection{Strain sequence}

\begin{figure}[!hbp]
\begin{center}
\includegraphics[scale=0.5, angle=0]{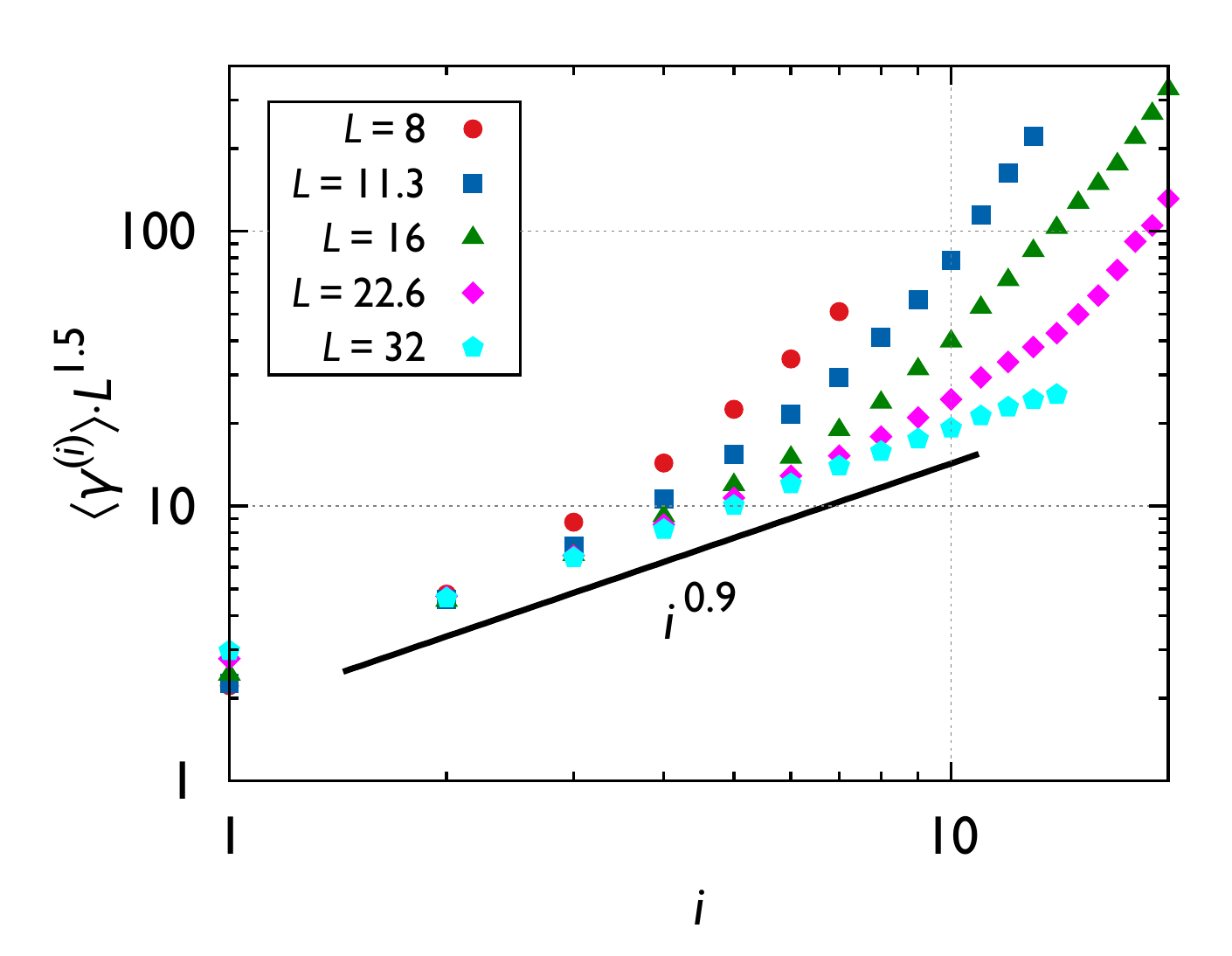}
\begin{picture}(0,0)
 \put(-205,144){\textsf{(a)}}
\end{picture}
\includegraphics[scale=0.5, angle=0]{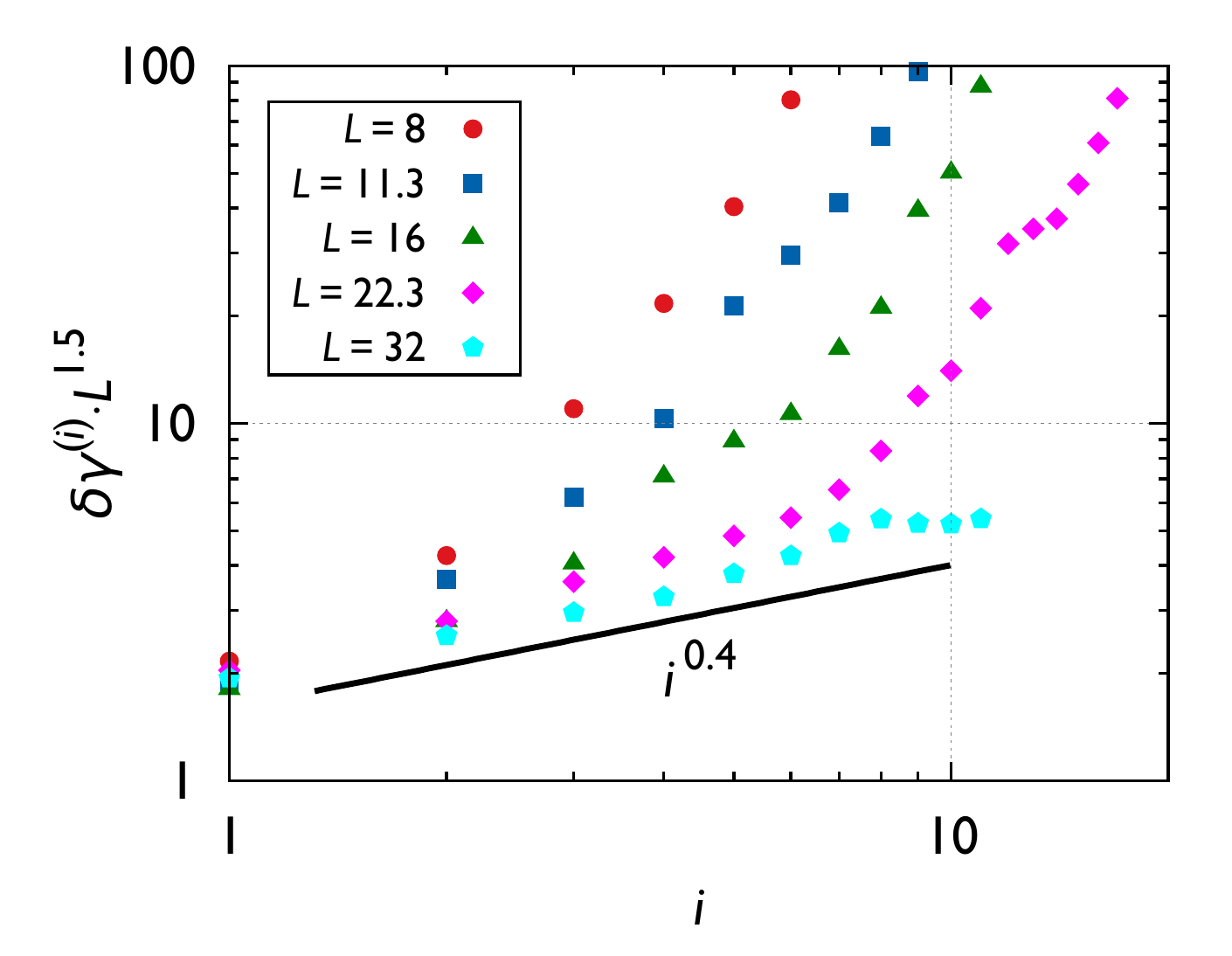}
\begin{picture}(0,0)
 \put(-205,144){\textsf{(b)}}
\end{picture}
\caption{\label{fig:gamma_i_tcddd} Statistics of the strain sequence $\gamma^{(i)}$ for TCDDD simulations of different system sizes. The average [panel (a)] and STD [panel (b)] of the plastic strain $\gamma^{(i)}$ at the \textit{i}th avalanche for different system sizes. The data are consistent with Eqs.~(\ref{eq:strain_sequence_average}) and (\ref{eq:strain_sequence_std}) (solid lines) if $i\gtrsim3$ and $\langle \gamma^{(i)}\rangle \lesssim 0.2$ with $\zeta = 0.9 \pm 0.05$ and $\xi = 1.5 \pm 0.1$.}
\end{center}
\end{figure}

Figure \ref{fig:gamma_i_tcddd} plots the average ($\langle \gamma^{(i)} \rangle$) and STD ($\delta \gamma^{(i)}$) of the strain sequence obtained for different system sizes. The curves are consistent with Eqs.~(\ref{eq:strain_sequence_average}) and (\ref{eq:strain_sequence_std}) found for the SCPM, with exponents $\zeta = 0.9 \pm 0.05$ and $\xi=1.5 \pm 0.1$

The overview of the introduced exponents, and their measured values are summarized in Table \ref{tab:example}.

\renewcommand{\arraystretch}{1.4}
\begin{table*}
\caption{\label{tab:example}Summary of exponents used in the paper.}
\begin{ruledtabular}
\begin{tabular}{L{1.5cm} L{6.5cm} C{2.5cm} C{2.2cm} C{2.2cm} C{2.2cm}}
Exponent & Description & Value predicted by theory & Value for SCPM & Value for TCDDD & Value for CADDD\\
\hline
$\nu$ & Characterizes threshold stress distribution, see Eq.~(\ref{eqn:link_distrib}) & - & $1.0$, $1.4$, and $2.0$\footnotemark[1] & $1.4 \pm 0.05$ & - \\
$\eta$ & Describes the relation between the system size $L$ and the total number of links $M$, see Eq.~(\ref{eq:exponent_eta_def}) & - & $2.0 \pm 0.05$ & $1.6 \pm 0.1$ & - \\
$\zeta$ & Characterizes the strain sequence, see Eq.~(\ref{eq:gamma_i}) & - & $1.0 \pm 0.05$ & $0.9 \pm 0.05$ & - \\
$\xi$ & Characterizes the system size dependence of the average avalanche size, see Eq.~(\ref{eq:av_avalanche_size}) & $\eta \zeta$ & $2.0 \pm 0.1$ & $1.5 \pm 0.1$ & - \\
$\alpha$ & Exponent of the power-law characterizing the microplastic regime of the stress-strain curves, see Eq.~(\ref{eq:avg_stress_strain}) & $(\nu \zeta)^{-1}$ & $1.0\pm 0.05$, $0.7\pm 0.05$, and $0.5 \pm 0.05$ & $0.8\pm 0.05$ & $0.8 \pm 0.05$\\
$\beta$ & Exponent characterizing the system size dependence of the stress fluctuations, see Eq.~(\ref{eq:stress_fluct}) & $\eta/2$ & $1.0 \pm 0.05$ &  $0.8 \pm 0.05$ & $0.8 \pm 0.05$ \\
\end{tabular}
\end{ruledtabular}
\footnotetext[1]{This exponent is an input parameter for the SCPM model}
\end{table*}
\renewcommand{\arraystretch}{1.0}

\section{Plasticity model based on extreme statistics}
\label{sec:model}

In this section a simple model for stochastic plasticity is introduced for stress-controlled loading. In this case the stress-strain curves are step-like and can be characterized by the stress and strain values $\tau^{(i)}$ and $\gamma^{(i)}$ corresponding to each step (see the sketch in Fig.~\ref{fig:stress_strain_sketch}). In the following, we propose assumptions for the stress and strain sequence and then combine them to obtain statistical predictions for the stress-strain curves. As we shall see, the proposed scaling forms will be identical to those obtained numerically in the previous section, so, for clarity, the same notation will be used for the exponents as before (see Table \ref{tab:example}).

\subsection{Stress sequence}
\label{sec:stress_seq}

Recently Derlet and Maa\ss  \ proposed a probabilistic approach to explain size effects observed in crystalline specimens \cite{derlet2015probabilistic}. They assumed, in accordance with the main idea of the SCPM, that plasticity occurs via irreversible structural excitations and that the material inhomogeneities are represented by a critical stress distribution $P(\tau) = \frac{\rmd}{\rmd \tau} \Phi(\tau)$, with $\Phi(\tau)$ being the cumulative distribution. Since during stress increase the weakest sites are activated, only the $\tau \to 0$ asymptote of this distribution is important, for which they used a power-law form
\begin{equation}
\Phi(\tau) \approx \left(\frac{\tau}{\tau_0} \right)^{\nu},\quad\text{if }\tau \to 0
\label{eqn:link_distrib}
\end{equation}
with $\nu\ge1$. It was also assumed, that the subsequent events are independent, so the spatial correlations of stress and plastic strain present in the SCPM were completely neglected. In this case, if the sample consists of $M$ sites where plastic events may occur, then the $i$th stress value $\tau^{(i)}$ in the $M \to \infty$ case follows Weibull order statistics \cite{rinne2008weibull}. In particular, the first event $\tau^{(1)}$ is Weibull distributed with shape parameter $\nu$:
\begin{equation}
\Phi^{(1)}(\tau^{(1)}) = 1 -\exp\left(-\frac1M \left(\frac{\tau^{(1)}}{\tau_0}\right)^\nu \right),
\label{eqn:weibull}
\end{equation}
the expected value for $\tau^{(i)}$ is
\begin{equation}
\left \langle \tau^{(i)} \right \rangle = \frac{\tau_0}{M^{1/\nu}} \frac{\Gamma\left(i+\frac{1}{\nu}\right)}{\Gamma(i)} \approx \tau_0 \left( \frac{i}{M} \right)^{\frac{1}{\nu}},
\label{eq:tau_i}
\end{equation}
and for the standard deviation of $\tau^{(i)}$ one obtains
\begin{equation}
\begin{split}
%\delta \tau^{(i)} &= \left\{ \left( \frac{\tau_0}{M^{1/\nu}}\right)^2 \left[ \frac{\Gamma\left(i+\frac{2}{\nu}\right)}{\Gamma(i)} - \left( \frac{\Gamma\left(i+\frac{1}{\nu}\right)}{\Gamma(i)} \right)^2 \right] \right\}^{1/2}\\
\delta \tau^{(i)} &= \sqrt{ \left( \frac{\tau_0}{M^{1/\nu}}\right)^2 \left[ \frac{\Gamma\left(i+\frac{2}{\nu}\right)}{\Gamma(i)} - \left( \frac{\Gamma\left(i+\frac{1}{\nu}\right)}{\Gamma(i)} \right)^2 \right] }\\
&\approx \frac{\tau_0}{i^{1/2}} \left( \frac{i}{M} \right)^{\frac{1}{\nu}}.
\end{split}
\label{eq:delta_tau_i}
\end{equation}
This means that the relative fluctuation decreases as $\delta \tau^{(i)}/\left \langle \tau^{(i)} \right \rangle \approx i^{-1/2}$, independent of the number of sites $M$. We also note, that the error of these approximations is less than 1\% if $i \gtrsim5$ and $M \gg i$, and that the distribution of $\tau^{(i)}$ tends to normal for large $i$ values.

Finally, one has to find the relation between $M$ and the linear system size $L$. It is natural to assume, that the activation sites are homogeneously distributed, and that their density does not depend on the sample size. This indicates $M\propto L^d$, with $d$ the dimension of the system. As we shall see below this hypothesis must be refined for DDD systems due to anomalous system size scaling. Therefore, exponent $\eta$ is introduced as:
\begin{equation} \label{eq:exponent_eta_def}
M\propto L^\eta,
\end{equation}
leading to
\begin{equation}
\Phi^{(1)}(\tau^{(1)}) = 1 -\exp\left(-\left(\frac 1{\tau_0}\frac{\tau^{(1)}}{L^{\eta/\nu}}\right)^\nu \right),
\end{equation}
\begin{equation}
\left \langle \tau^{(i)} \right \rangle \approx \tau_0 L^{-\eta/\nu} i^{1/\nu},
\label{eq:tau_i_2}
\end{equation}
and
\begin{equation}
\delta \tau^{(i)} \approx \tau_0 L^{-\eta/\nu} i^{1/\nu-1/2}.
\label{eq:delta_tau_i_2}
\end{equation}

%As said, this model assumes that the subsequent events are independent and that an event does not affect the critical stress of the other sites. In the DDD SCPM models these assumptions are obviously not valid, due to the internal stress redistributions. Surprisingly, it will be seen, that Eq

\subsection{Strain sequence}
\label{sec:strain_seq}

According to numerous recent experimental and numerical studies, the plastic strain increments, corresponding to the strain burst events, exhibit power-law distribution \cite{miguel2001intermittent, dimiduk2006scale, zaiser2008strain, ispanovity2014avalanches}. However, the scale-free behavior is observed only in a bounded region since (i)
%
%It is, however, chopped of both at small and large scales, and the upper cutoff diverges only at criticality and infinite sample sizes. This means that for finite %specimens and small stresses the support of the strain burst size distribution is bounded from both sides 
%
%However, this scale free distribution is observed only in a finite, though typically quite wide, region, bounded from both directions.
%
at large strain jumps the distribution is chopped off due to finite system size, and (ii) in the case of very small strain bursts, deviation from the power-law is necessary otherwise the strain burst distribution could not be normalized. The physical origin of this lower cutoff is that here individual dislocation motion dominates over collective dislocation dynamics. In summary, the strain burst size ($\Delta \gamma$) distribution looks as
\begin{equation}
	P_\text{sb}(\Delta \gamma) = C s^{-\tau_a} f(\Delta \gamma/\Delta \gamma_u), \text{ if }\Delta \gamma > \Delta \gamma_l,
\end{equation}
where $\Delta \gamma_l$ and $\Delta \gamma_u$ represent the lower and upper cutoff, respectively, $\tau_a$ is the avalanche size exponent, $C$ is a normalization factor, and $f$ is the cutoff function that decays faster than algebraically for large arguments and $f(x) \to 1$, if $x \to 0$.

It follows, that for finite system sizes and small applied stresses, due to the cutoffs, distribution $P_\text{sb}(\Delta \gamma)$ has finite moments, in particular, finite mean and variance. The recent numerical study of 2D DDD systems of Ispánovity \emph{et al.}\ showed that in the microplastic regime $\tau_a \approx 1$, the upper cutoff $\Delta \gamma_u$ depends weakly on the applied stress, and it exhibits anomalous system size dependence.\cite{ispanovity2014avalanches} Consequently, the mean and variance of strain increment can be written as
\begin{eqnarray}
\langle \Delta \gamma \rangle = \frac{s_0}{L^\xi}, \label{eq:av_avalanche_size}\\
\delta (\Delta \gamma) = \frac{s_1}{L^\xi},
\end{eqnarray}
respectively, where $\xi$ is the exponent characterizing the system size dependence of the avalanche sizes, and $s_0$ and $s_1$ are appropriate constants, that may depend on the applied stress. Here $\xi=2$ corresponds to normal scaling, where the total plastic slip during an avalanche is independent of the system size, whereas $\xi<2$ indicates anomalous scaling.

In order to derive predictions for the strain sequence it is assumed that the size of subsequent strain bursts is uncorrelated. Then from the central limit theorem it follows, that for $i \gg 1$ and small applied stresses $\gamma^{(i)}$ is distributed normally, and
\begin{gather}
\langle \gamma^{(i)} \rangle = i^\zeta \langle \Delta \gamma \rangle = i^\zeta \frac{s_0}{L^\xi}, \label{eq:gamma_i}\\
\delta \gamma^{(i)} = i^{\zeta-0.5} \delta (\Delta \gamma) = i^{\zeta-0.5} \frac{s_1}{L^\xi}. \label{eq:delta_gamma_i},
\end{gather}
with $\zeta = 1$.

\subsection{Stress-strain curves}
\label{sec:stress_strain_curves}

Since the sequences $\tau^{(i)}$ and $\gamma^{(i)}$ give a full description of the stress-strain curve, in the following the expressions for $\tau^{(i)}$ and $\gamma^{(i)}$ derived above are combined to obtain statistical properties of the stress-strain curves. It was predicted that both $\gamma^{(i)}$ and $\tau^{(i)}$ are distributed normally [see Eqs.~(\ref{eq:gamma_i}, \ref{eq:delta_gamma_i}) and Eqs.~(\ref{eq:tau_i},\ref{eq:delta_tau_i})] for $i\gg 1$. By ``inverting'' $\gamma^{(i)}$ to express $i$ at a given plastic strain $\gamma$, and then inserting $i(\gamma)$ into $\tau^{(i)}$ one obtains that for $i\gg 1$ $\tau$ is distributed normally and:
\begin{gather}
\langle \tau \rangle = \frac{\tau_0}{s_0^{1/\nu \zeta}} L^{(\xi/\zeta-\eta)/\nu} \gamma^{1/\nu \zeta}, \label{eq:tau}\\
\delta \tau = \tau_2 L^{(\xi/\zeta-\eta)/\nu - \xi/2 \zeta} \gamma^{1/\nu\zeta - 1/2\zeta}. \label{eq:delta_tau}
\end{gather}
This means that the average stress-strain curve starts as a power-law and if $\xi/\zeta \ne \eta$ then it has a system size dependence even at very large system sizes. To exclude this nonphysical situation one requires
\begin{equation}
\xi = \eta \zeta.
\label{eq:size_eff}
\end{equation}
In this case the average and the fluctuation of the stress-strain curve behave as
\begin{gather}
\langle \tau \rangle \propto \gamma^{\alpha}, \label{eq:tau2} \\
\delta \tau \propto L^{-\beta},
\label{eq:delta_tau_2}
\end{gather}
with
\begin{gather}
\alpha = 1/\nu \zeta, \\
\beta = \eta/2
\end{gather}
Since $\eta$ is positive, this means that stress fluctuations decrease as size increases, that is, one obtains a smooth stress-strain curve for very large samples, as expected.

To summarize this section, a weakest ling assumption proposed by Derlet and Maa\ss \ was adopted for the stress sequence [Eqs.~(\ref{eq:tau_i}, \ref{eq:delta_tau_i})],\cite{derlet2015probabilistic} and a straightforward rule for the strain sequence was proposed which is able to capture the anomalous system size dependence of 2D DDD systems [Eqs.~(\ref{eq:gamma_i}, \ref{eq:delta_gamma_i})].\cite{ispanovity2014avalanches} The combination of the two series has led to statistical predictions on the stress-strain curves [Eqs.~(\ref{eq:tau}, \ref{eq:delta_tau})], which, in fact, coincide with the numerical findings described in Sec.~\ref{sec:results}.

\section{Discussion}

In the preceding two sections numerical results and a theory were presented yielding identical scaling forms for the average and fluctuation of the individual stress-strain curves and the stress/strain sequence in the microplastic regime. The exponents introduced to describe these quantities and their measured/predicted values are summarized in Table \ref{tab:example}. In the following discussion we highlight the most important consequences of these findings.

Firstly, we consider the main idea of SCPM that the material can be decomposed into local units each characterized by a yield threshold. This non-trivial assumption implies that the distribution of $\tau^{(1)}$, i.e., the stress at the onset of the first event, must follow a weakest link distribution, so, the fact that for DDD a Weibull distribution was found to describe $P(\tau^{(1)})$ supports this fundamental hypothesis.  In addition, it provides us access to the individual link distribution for dislocation structures, since the shape parameter of a Weibull $\nu$ unambiguously determines the asymptote of the underlying link distribution, in this case a power-law of Eq.~(\ref{eqn:link_distrib}). As such, exponent $\nu$ emerges as a central parameter that also influences the power-law exponent of the plastic stress-strain relation ($\alpha=1/\nu\zeta$), that is, the amount of plasticity in the microplastic regime (see Fig.~\ref{fig:avg_stress_strain_SCPM}). The origin of $\nu=1.4$ for DDD systems is not addressed in this paper, it may be influenced by the internal structure of dislocations, like slip systems, patterns, etc. It is noted, however, that a similar analysis of the average stress-strain curves performed earlier on 3D DDD simulations and micropillar compressions yielded $\alpha \approx 0.8$ in both cases,\cite{ispanovity2010submicron,Ispanovity2013} hinting at some generality in the value of $\nu$ (with the straightforward assumption of $\zeta \approx 1$).

\begin{figure}[!htbp]
\begin{center}
\includegraphics[scale=0.5]{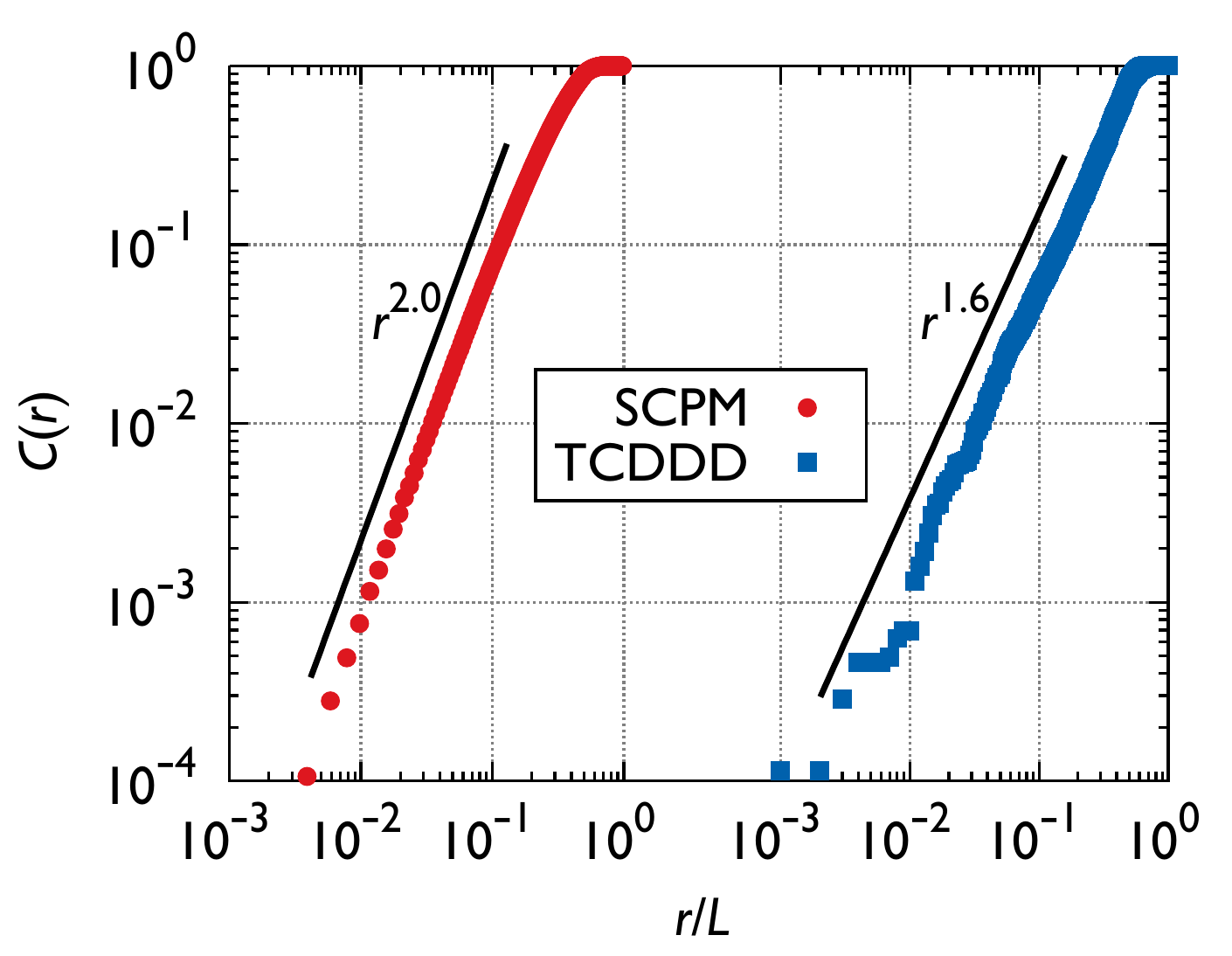}
\caption{\label{fig:corr_int} Correlation integral of the avalanche positions for the SCPM and TCDDD models. The measured data are consistent with $C(r)\propto r^\eta$, the latter indicated by the solid lines.}
\end{center}
\end{figure}

Although the behavior of the stress sequence shows strong similarity between DDD and SCPM, exponent $\eta$ characterizing the system size dependence of the number of ,,links" of the system $M$ differs considerably. For SCPM $\eta \approx 2$ was found, that corresponds to proportionality between $M$ and the 2D system size, whereas for DDD a significantly smaller value of $\eta \approx 1.5$ was obtained, hinting at a fractal-like structure of the weakest regions of the system. In order to quantify this conjecture we consider the correlation integral $C(r)$ of the initiation points of the events, defined as the probability of the distance of two such arbitrarily chosen points being smaller than $r$. A fractal dimension $d$ can be defined from the asymptotic behavior as $C(r) \propto r^d$. Indeed, according to Fig.~\ref{fig:corr_int} the $C(r)\propto r^\eta$ is a very good approximation both for SCPM and DDD. Although the explanation of this difference is out of the scope of the present paper we mention that the fact that plasticity accumulates on a fractal-like sub-domain of the system may explain the recent findings on the long-range nature of dynamical correlations and peculiar critical behavior of 2D DDD systems.\cite{ispanovity2014avalanches} In addition, it echoes on the experimental findings of Weiss et al., where it was found that during creep deformation of an ice single crystal AE signals initiate from a fractal sub-volume of the specimen with dimension $\sim$2.5.\cite{weiss2003three}

\begin{figure}[!htbp]
\begin{center}
\includegraphics[scale=0.5]{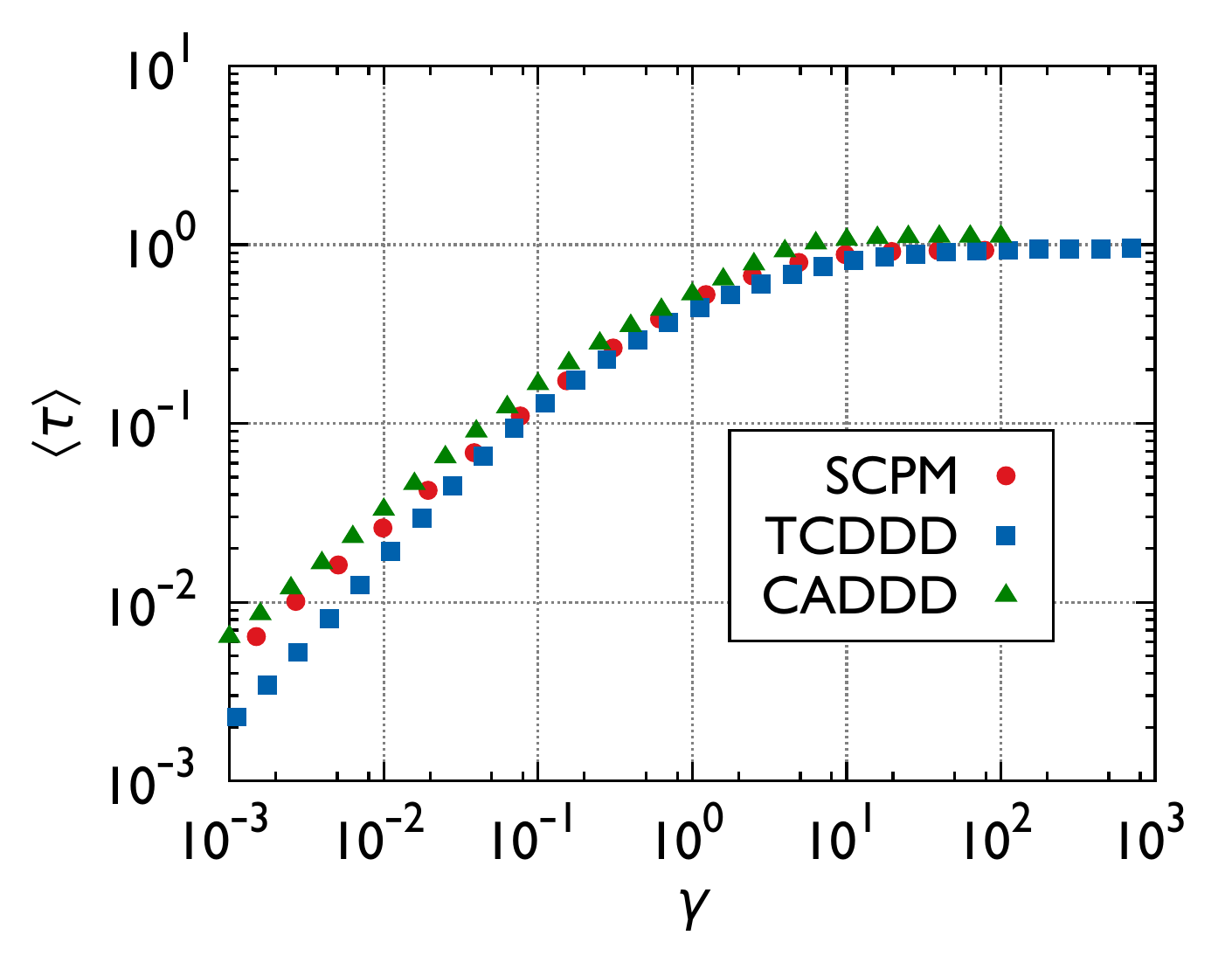}
\caption{\label{fig:comparison} Stress-strain curves obtained by the three different plasticity models. For the SCPM $\nu=1.4$, $\Delta \gamma^\text{pl} = 6$, and $\tau_w = 2$ was chosen.}
\end{center}
\end{figure}

The plastic response of micron-scale samples usually show strong size-effects, for instance, the strength of micro-pillars and nano-pillars may increase two orders of magnitude by reducing their size. The models of this paper, on the other hand, only exhibit very weak size-effects with a maximum increase of $\sim$10\% in the measured average stress (see Figs.~\ref{fig:avg_stress_strain_SCPM}(a) and \ref{fig:avg_stress_strain_DDD}). One thus concludes, that the strong stress increase observed in experiments is not due to the finiteness of the dislocation content, but is the result of the free surfaces, where a significant fraction of dislocations can leave the crystal.

In summary, the two models behave similarly in the microplastic regime, where the simple plasticity theory introduced that neglects internal correlations and the spatial extent of avalanches is able to properly describe the fluctuating plastic response. More is true, however, since the average and scatter of the stress-strain curves obtained by SCPM and DDD are quite similar for moderately large strains, too, where the plasticity model is clearly not applicable due to strong internal correlations (that is, in the $0.1\lesssim \gamma \lesssim 10$ region in Fig.~\ref{fig:comparison}). So, it seems that SCPM can correctly capture the internal stress and strain correlations developing upon increasing strain. At very large deformations ($\gamma \gtrsim10$), however, specific dislocation patterns develop in the DDD models with characteristic scales comparable to the system size,\cite{zhou2015dynamic,PhysRevB.91.054106} the SCPM cannot account for. Therefore, similarity between the models is not expected in this regime.

\section{Summary}

In this paper we have demonstrated that the SCPM model introduced earlier is able to quantitatively describe the stochastic properties of crystalline plasticity. Using a simple theoretical model for the microplastic regime based on the subsequent activation of the weakest links in the sample we derived a method how to calibrate the parameters of the SCPM based on lower-level DDD simulations. The proposed methodology does not only represent a bridge between micro- and meso-scales, but also gives insight into the nature of stochastic processes characterizing plasticity. The current paper has focused on crystal plasticity and a simple 2D DDD representation, but the authors do not see any reason why the proposed plasticity model and the multi-scale methodology would not be applicable for more involved DDD models or amorphous materials. The verification of this conjecture is delegated to future work, and is expected to open new perspectives in the applicability of stochastic continuum plasticity models.

\section*{Acknowledgments}

PDI thanks Peter Derlet for fruitful discussions. Financial supports of the Hungarian Scientific Research Fund (OTKA) under
contract numbers K-105335  and PD-105256 and of the European Commission under grant agreement No.
CIG-321842 are also acknowledged. PDI is also supported by the J\'anos Bolyai Scholarship of the
Hungarian Academy of Sciences.

\bibliography{multiscale_ddd_ca}

\end{document}